\documentclass[%
reprint,
showpacs,
amsmath,amssymb, aps,
prb,
floatfix
]{revtex4-1}

\usepackage{graphicx}% Include figure files
\usepackage{dcolumn}% Align table columns on decimal point
\usepackage{bm}% bold math
\usepackage{hyperref}
\usepackage{xcolor}
\hypersetup{colorlinks,linkcolor={red},citecolor={blue},urlcolor={blue}}

\def\be{\begin{equation}}
\def\ee{\end{equation}}
\def\bea{\begin{eqnarray}}
\def\eea{\end{eqnarray}}

\def\ra{\rangle}
\def\la{\langle}
\def\s{\sigma}

\def\ve{\varepsilon}

\usepackage{color}
\definecolor{DarkGreen}{rgb}{0,0.6,0}

\newcommand{\dg}[1]{{#1}^\dagger}

\newcommand{\ket}[1]{\left|{#1}\right\rangle}

\begin{document}

\title{Topological and non-topological features of generalized Su-Schrieffer-Heeger models}

\author{N. Ahmadi}
\affiliation{Department of Physics, Institute for Advanced Studies
	in Basic Sciences (IASBS), Zanjan 45137-66731, Iran}

\author{J. Abouie}
\email[]{jahan@iasbs.ac.ir}
\affiliation{Department of Physics,
	Institute for Advanced Studies in Basic Sciences (IASBS), Zanjan
	45137-66731, Iran}

\author{D. Baeriswyl}
\affiliation{Department of Physics, University of Fribourg, CH-1700 Fribourg, Switzerland}
\begin{abstract}

The (one-dimensional) Su-Schrieffer-Heeger Hamiltonian, augmented by spin-orbit coupling and longer-range hopping, is studied at half filling for an even number of sites. The ground-state phase diagram depends sensitively on the symmetry of the model. Charge-conjugation (particle-hole) symmetry is conserved if hopping is only allowed between the two sublattices of even and odd sites. In this case (of BDI symmetry) we find a variety of topologically non-trivial phases, characterized by different numbers of edge states (or, equivalently, different quantized Zak phases). The transitions between these phases are clearly signalled by the entanglement entropy. 
Charge-conjugation symmetry is broken if hopping within the sublattices is admitted (driving the system into the AI symmetry class). We study specifically next-nearest-neighbor hopping with amplitudes $t_a$ and $t_b$ for the $A$ and $B$ sublattices, respectively. For $t_a=t_b$ parity is conserved, and also the quantized Zak phases remain unchanged in the gapped regions of the phase diagram. However, metallic patches appear due to the overlap between conduction and valence bands in some regions of parameter space. The case of alternating next-nearest neighbor hopping, $t_a=-t_b$, is also remarkable, as it breaks both charge-conjugation $C$ and parity $P$ but conserves the product $CP$. Both the Zak phase and the entanglement spectrum still provide relevant information, in particular about the broken parity. Thus the Zak phase for small values of $t_a$ measures the disparity between bond strengths on $A$ and $B$ sublattices, in close analogy to the proportionality between the Zak phase and the polarization in the case of the related Aubry-Andr\'e model.
\end{abstract}

\date{\today}

%\pacs{}

\maketitle

%%%%%%%%%%%%%%%%%%%%%%%%%%%%%%%%%%%%%%%%%%%%%%% Introduction %%%%%%%%%%%%%%%%%%%%%%%%%%%%%%%%%%%%%%%%%%%%%%%%%%%
\section{Introduction}\label{sec:intro}
Since the discovery of quantum Hall effects,\cite{PhysRevLett.45.494,PhysRevLett.48.1559} it has been realized that many materials experience different orders in which the topology plays a vital role.
Theoretical and experimental investigation of topological states became subsequently an active field of research in different disciplines, such as condensed matter physics,\cite{Rach_002018,RevModPhys.82.3045} photonics,\cite{LEHUR2016808} and ultracold atomic gases.\cite{NAture} 
Recently, much attention has been devoted to the identification and classification of different topological phases of matter.
It is quite well known that symmetry and dimension play key roles in the classification of topological properties.\cite{PhysRevX.8.031079,RevModPhys.88.035005} For one-dimensional (1D) Hermitian systems with
chiral symmetry, the topological properties can be characterized by the quantized Berry phase across the first Brillouin zone (Zak phase).\cite{PhysRevLett.62.2747}
In the absence of chiral symmetry, the Berry phase is not quantized, and it is not a topological invariant. 

A remarkable quantum-mechanical phenomenon is entanglement. Aside from its notable applications in quantum
computation\cite{nielsen00}, it
can be used to probe different phase transitions\cite{osterlo} as
well as topological properties of many-body quantum states. For
instance, the topological entanglement entropy, the most commonly used measurement of
entanglement, is directly related to the total quantum dimension of
fractional
quasi-particles.\cite{PhysRevLett.96.110404}
Moreover, Li and Haldane showed that the eigenvalues of the
reduced density matrix, called entanglement spectrum,
contains complete information about different phases and  various
phase transitions.\cite{PhysRevLett.101.010504} The notion of the
entanglement spectrum has led to novel insights in the physics of
quantum Hall systems\cite{PhysRevLett.101.010504,PhysRevLett.104.156404,PhysRevB.85.125308},
quantum spin systems in one
\cite{PhysRevB.81.064439,PhysRevLett.105.077202,PhysRevLett.105.116805,PhysRevB.85.075125} and two
\cite{PhysRevLett.105.080501,PhysRevB.96.121115} dimensions,
topological
insulators\cite{PhysRevLett.104.130502,PhysRevB.88.115114} and
bilayer
lattices.\cite{PhysRevB.85.054403,1367-2630-15-5-053017,1742-5468-2016-11-113101}

One of the most simple topologically non-trivial systems is the SSH model, introduced by Su, Schrieffer and Heeger in the late seventies \cite{PhysRevLett.42.1698} to describe both elastic and electronic properties of polyacetylene, a chain of CH groups. In this model the elastic and electronic degrees of freedom interact because relative displacements of neighbouring CH groups change the overlap of $\pi$-electron orbitals. For a density of one $\pi$-electron per CH group, the polymer chain is dimerized in the ground state, i.e., it exhibits an alternating sequence of bond lengths. The dimerization leads to an electronic energy gap between a filled valence band and an empty conduction band. There are two distinct dimerization patterns, and geometric constraints (or doping) can force the coexistence of the two sequences with domain walls separating them. Thus a ring with an odd number of sites has necessarily a domain wall (soliton) in the ground state at half filling. These ``intrinsic defects'' are also referred to as topological solitons, in view of the origin of their stability. Remarkably, a domain wall generates a localized wave function with an energy at midgap. In the context of quantum field theory the existence of such states due to the coupling of fermions to a classical field with a kink-like spatial dependence has been discovered already before the advent of the SSH model \cite{Jackiw_76}. In quantum chemistry, such states have been proposed even earlier in the form of bond-alternation defects \cite{Pople_62}. Experiments on polyacetylene indicate that these solitons play an important role in the electronic and optical properties of the material \cite{Heeger_88}. 

In recent years the electronic part of the SSH Hamiltonian has been frequently used as a toy model for topological insulators. Bond alternation is taken into account by assuming alternating hopping integrals $t(1\pm \delta)$ with a site-independent dimerization parameter $\delta$ $(\vert\delta\vert<1)$. The Hamiltonian reads
\begin{equation}\label{Eq:ham_SSH}
H_{\mbox{\scriptsize SSH}}=t\sum_{i\sigma}\left[ \left( 1+\delta \right) a_{i\s}^{\dagger}b_{i\s}^{\phantom{}}
+ (1-\delta )a_{i+1\s}^{\dagger}b_{i\s}^{\phantom{}} + \mbox{h.c.}\right]\, ,
\end{equation}
where $\dg a_{i,\s} (a_{i,\s})$ and $\dg b_{i,\s} (b_{i,\s})$ are fermion
creation (annihilation) operators with spin $\s$
on the sublattices A and B, respectively, in the unit cell $i$ ($i=1,...,L$). We have implicitly assumed an even number of sites ($2L$). For periodic boundary conditions, the energy spectra for the two dimerization patterns ($\delta<0$ and $\delta>0$) are identical, but the eigenstates produce different Berry (or Zak) phases \cite{Delplace_11}. For open boundary conditions, the two cases have distinct energy spectra. For $\delta<0$, i.e., for two weak bonds at the chain ends, there are two edge states with levels close to midgap. This is the topologically nontrivial phase. For $\delta>0$, i.e., for two strong bonds at the chain ends, there are no edge states.

It is important to notice that the dimerization parameter $\delta$ may be the result of a spontaneous symmetry breaking, as in polyacetylene\cite{Heeger_88}, or it may be produced by external fields, as for cold atoms in optical lattices \cite{nature11, Cooper_19}. In the latter case, $\delta$ is fixed and cannot be adjusted to lower the energy. However, in the former case the dimerization pattern can change. In such a situation, the topological phase with two dangling bonds at the chain ends ($\delta<0$) is unstable, because it has a higher energy than the ground-state configuration for $\delta>0$. In this paper we assume generally $\delta>0$, i.e. we study the case for which the ground state of the SSH Hamiltonian is topologically trivial. Therefore the topologically non-trivial phases will be produced by extensions of the model.

There are plenty of quasi-one-dimensional conductors, such as the family of organic charge-transfer salts, where the coupling between electrons and elastic degrees of freedom is not produced by the bond-length dependence of the hopping term, but by shifts of local electronic energies due to
intra-molecular (on-site) displacements (Holstein model \cite{Holstein_59}). In such a case, the Peierls instability leads to an alternating charge density linked to an energy gap $2\Delta$. If we again neglect the lattice degrees of freedom, we arrive at the Aubry-Andr\'e Hamiltonian (for the special case of period-two on-site energies)\cite{Aubry_80}
\begin{eqnarray}\label{Eq:ham_CDW}
H_{\mbox{\scriptsize AA}}&=\sum_{i\sigma}\left[
t\left( a_{i\s}^{\dagger}b_{i\s}^{\phantom{}}+ a_{i+1\s}^{\dagger}b_{i\s}^{\phantom{}}+ \mbox{h.c.}\right)\right.\nonumber\\
&+\left.\Delta\left(a_{i\s}^{\dagger} a_{i\s}^{\phantom{}}-b_{i\s}^{\dagger} b_{i\s}^{\phantom{}}\right)\right]\, .
\end{eqnarray}

The Hamiltonians (\ref{Eq:ham_SSH}) and  (\ref{Eq:ham_CDW}) have both different symmetries and different topological properties. However, a simple transformation brings Eq. (\ref{Eq:ham_CDW}) into the form of Eq. (\ref{Eq:ham_SSH}), augmented by second- and third-neighbor hopping. This is shown in Appendix \ref{sec:CDW-SSH}. A combination of Eqs. (\ref{Eq:ham_SSH}) and  (\ref{Eq:ham_CDW}) leads to the Rice-Mele Hamiltonian \cite{Rice_82}, which can also be transformed to a generalized SSH model.

Some recent studies have already considered the effects of longer-range hopping. Thus it was found that new phases can be produced by adding next-nearest neighbor hopping to Eq. (\ref{Eq:ham_SSH}) \cite{PhysRevB.89.085111}. There is an important difference between ``odd hoppings'' (connecting only sites of different sublattices) and ``even hoppings'' (connecting sites within the sublattices), as they lead to different symmetry classes \cite{PhysRevB.99.035146}. In the presence of exclusively odd hoppings the system belongs to the BDI symmetry class (Altland-Zirnbauer classification \cite{PhysRevB.55.1142,RevModPhys.88.035005}), where the topology is characterized by a quantized Berry phase.

Another extension of the Hamiltonian (\ref{Eq:ham_SSH}) concerns
spin-orbit coupling (SOC), which modifies substantially the band structure and leads to new topological phases \cite{Vayrynen_11, 0295-5075-107-4-47007, PhysRevB.94.125119, PhysRevB.96.205424}. Experimentally, effects of a synthetic SOC on the edge modes of the SSH model have been investigated using a photonic system \cite{Whittaker_19}.

In this paper we consider a generalized SSH model by adding both spin-orbit interaction and longer-range hopping to Eq. (\ref{Eq:ham_SSH}). In the first part, we treat the case with only odd hopping. Several topological states are found, which can be characterized by quantized Zak phases or, equivalently, by the number of edge states. We also discuss the critical behavior of the topological transitions separating the different phases. 
The energy levels of the entanglement spectrum, the entanglement gap and the entanglement entropy correctly reproduce the ground state phase diagram and help to identify different topological phases. In the second part, we show that the addition of hopping between next-nearest neighbors has drastic effects. Besides putting the system into a different symmetry class, this term strongly affects the phase transitions, the Zak phase and entanglement properties. Even if the ground states can no longer be labeled by integer winding numbers, interesting relations between the Zak phases and local quantities appear (polarization for the Aubry-Andr\'e model, specific correlation functions for the generalized SSH Hamiltonian). We restrict ourselves to specific spatial symmetries, on the one hand to the case of conserved parity $P$ (where the next-nearest-neighbor hoppings are equal on the two sublattices), on the other hand to the case where parity is not conserved, but the product of charge conjugation $C$ and parity $P$ is (this happens if the next-nearest-neighbor hoppings have the same strengths but opposite signs). Entanglement properties remain relevant also for these ``topologically trivial'' systems, in particular for the $CP$-symmetric case.

The paper is organized as follows. Section \ref{sec:SO-SSH+Odd} deals with the chirally symmetric extended SSH model, where even hopping is excluded. The phase diagram of this model is presented in \ref{sec:pd}, entanglement is discussed in \ref{sec:entanglement} and the critical behavior is briefly described in 
\ref{sec:critical}. Section \ref{sec:SO-SSH+Even} is concerned with the effects of next-nearest-neighbor hopping, which breaks charge-conjugation symmetry. The special cases of $P$ and $CP$ symmetries are analyzed in \ref{sec:parity} and \ref{sec:CP}, respectively. The relation between the Zak phase and local physical quantities is elucidated in \ref{sec:CDW} by means of the polarization in the closely related Aubry-Andr\'e model. The paper is summarized in Section \ref{sec:summary}.

%%%%%%%%%%%%%%%%%%%%%%%%%%%%%%%%%%%%%%%%%%%%%%%%%%
\section{Topological phases and entanglement for chiral symmetry}\label{sec:SO-SSH+Odd}
%%%%%%%%%%%%%%%%%%%%%%%%%%%%%%%%%%%%%%%%%%%%%%%%%

In this section we study a generalized 1D SSH model with SOC in the presence of odd hopping. 
The Hamiltonian is 
\begin{equation}\label{Eq:ODDHamiltonian}
H=H_{\mbox{\scriptsize SSH}}+H_{\mbox{\scriptsize SO}}+H_{\mbox{\scriptsize NNNN}}\, ,
\end{equation}
where $H_{\mbox{\scriptsize SSH}}$ is defined by Eq. (\ref{Eq:ham_SSH}), $H_{\mbox{\scriptsize SO}}$ is the spin-orbit coupling
\begin{equation}\label{Eq:ham_SOC}
H_{\mbox{\scriptsize SO}}=\sum_{i\sigma}\left[\lambda a_{i\s}^{\dagger}b_{i-\s}^{\phantom{}}- \lambda' a_{i+1\s}^{\dagger}b_{i-\s}^{\phantom{}} + \mbox{h.c.}\right]\, ,
\end{equation}
which causes spin flips during hopping,
and $H_{\mbox{\scriptsize NNNN}}$ stands for hopping between third neighbors,
\begin{equation}\label{Eq:ham_NNNN}
H_{\mbox{\scriptsize NNNN}}=\sum_{i\sigma}\left[t_{ab} a_{i\s}^{\dagger}b_{i+1\s}^{\phantom{}}+  t_{ba} b_{i\s}^{\dagger}a_{i+2\s}^{\phantom{}} + \mbox{h.c.}\right].
\end{equation}
We limit ourselves to the case of half filling, where the number of particles is equal to the number of sites.

%%%%%%%%%%%%%%%
\subsection{Phase diagram}\label{sec:pd}
%%%%%%%%%%%%%%%
For periodic boundary conditions, the Hamiltonian (\ref{Eq:ODDHamiltonian}) can be written in momentum space as
\begin{equation}\label{Eq:SO-HamiltonianK}
	H = \sum_{k}\dg\psi_k H_k\psi_k\, ,
\end{equation}
where
\begin{equation}\label{Eq:Bloch}
	H_k=
	\begin{pmatrix}
		0 & \ve_k & 0 & f_k\\
		\ve_k^* & 0 & f_k^* & 0 \\
		0 & f_k & 0& \ve_k \\
		f_k^* & 0 & \ve_k^* & 0
	\end{pmatrix}
\end{equation}
is the so-called Bloch Hamiltonian and $\dg\psi_k=\begin{pmatrix} \dg a_{k\uparrow} & \dg
b_{k\uparrow}&\dg a_{k\downarrow}&\dg b_{k\downarrow}
\end{pmatrix}$. Here $\ve_k=t[1+\delta+(1-\delta)e^{-ik}]+t_{ab} e^{ik}+t_{ba} e^{-2ik}$, $f_k=\lambda - \lambda' e^{-ik}$, and 
$\ve_k^*,f_k^*$ are complex conjugates of $\ve_k,f_k$. We choose $t$ as the unit of energy, $t=1$, and assume $0<\delta<1$.

The Hamiltonian (\ref{Eq:SO-HamiltonianK}) possesses time reversal ($T$), particle-hole ($C$) and chiral ($S$) symmetries and therefore belongs to the BDI symmetry class of symmetry-protected topological (SPT) states.\cite{PhysRevB.55.1142,Ryu2010}. For the Bloch Hamiltonian (\ref{Eq:Bloch}) the corresponding operators are
$C = \sigma_0\otimes\s_z \kappa$, $T= \s_x \kappa \otimes \sigma_0$, and $S=TC= \s_x \otimes \s_z$, where $\kappa$ produces complex conjugation, $\s_{\alpha}$ with $\alpha=x,y,z$ are the Pauli matrices and $\sigma_0$ is the identity matrix. The Bloch Hamiltonian transforms as \cite{RevModPhys.88.035005, Ryu2010}
\begin{eqnarray}
\nonumber CH(k)C^{-1} &=& -H(-k),\\
\nonumber T H(k)T^{-1} &=& H(-k),\\
SH(k) S^{-1}&=& -H(k).
\end{eqnarray}

The matrix (\ref{Eq:Bloch}) is easily diagonalized and has eigenvalues 
\begin{equation}
E_{1k}=-E_{2k}=\vert\varepsilon_k+f_k\vert,\quad E_{3k}=-E_{4k}=\vert\varepsilon_k-f_k\vert,
\end{equation}
symmetric with respect to zero. In general the spectrum has a gap, but for special parameter values there exist wave vectors for which some energy eigenvalues vanish. This occurs if $\varepsilon_k=f_k$ or 
$\varepsilon_k=-f_k$. We find solutions of these equations for $k=0$ or $k=\pi$ if
\begin{equation}\label{Eq:linear}
t_{ab}=\left\{\begin{array}{l}-2-t_{ba}\pm(\lambda-\lambda')\, ,\\2\delta+t_{ba}\pm(\lambda+\lambda')\, .\end{array}\right.
\end{equation}
The gap can also close for other wave vectors provided that $\vert\delta-1+t_{ab}\mp\lambda'\vert\le 2\vert t_{ba}\vert$. In this case the condition for gap closing is given by the quadratic equation
\begin{equation}\label{Eq:quadratic}
t_{ab}^2-t_{ba}^2+(\delta-1\mp\lambda')t_{ab}+(\delta+1\mp\lambda)t_{ba}=0\, .
\end{equation}

Eqs. (\ref{Eq:linear}) and (\ref{Eq:quadratic}) define the boundaries between different topological phases, which we characterize by the Zak phase \cite{PhysRevLett.62.2747}, the Berry phase \cite{10.2307/2397741,PhysRevB.76.045302} for Bloch bands,
\begin{equation}
\Phi=\sum_{E<0}\int_{-\pi}^{\pi}\left\langle u_k| i \partial _k
u_k\right\rangle dk,\label{Eq:Berry}
\end{equation}
where the sum is over the two occupied eigenstates $\ket{u_k}$ of the Bloch Hamiltonian. The Zak phase is not invariant with respect to spatial translations or gauge transformations \cite{nature11, Cooper_19}, but it can be defined in such a way that its values reflect the number of edge modes.\cite{Delplace_11} This bulk-boundary correspondence remains valid if one adds the spin-orbit term. Indeed, for the Hamiltonian $H=H_{\mbox{\scriptsize SSH}}+H_{\mbox{\scriptsize SO}}$ one finds
two (non-trivial) topological phases, one with $\Phi=\pi$ and one pair of edge modes, the other with $\Phi=2\pi$ and two pairs of edge modes. \cite{PhysRevB.94.125119}

The variety of phases increases even more if third-neighbor hopping is added. To calculate the various Zak phases for the full Hamiltonian 
(\ref{Eq:ODDHamiltonian}), we proceed as follows. The eigenvectors of $H_k$ for the negative energy eigenvalues are chosen as
\begin{eqnarray}\label{Eq:vectors}
\vert u_k\rangle=\frac{1}{2}\left(\begin{array}{c}1\\-e^{{-i\varphi_k^{(+)}}}\\1\\-e^{{-i\varphi_k^{(+)}}}\end{array}\right)\quad\mbox{for }E=-\vert\varepsilon_k+f_k\vert\, ,\nonumber\\
\vert u_k\rangle=\frac{1}{2}\left(\begin{array}{c}1\\-e^{{-i\varphi_k^{(-)}}}\\-1\\e^{{-i\varphi_k^{(-)}}}\end{array}\right)\quad\mbox{for }E=-\vert\varepsilon_k-f_k\vert\, ,
\end{eqnarray}
where
\begin{equation}
e^{i\varphi_k^{(\pm)}}= \frac{\varepsilon_k\pm f_k}{\vert\varepsilon_k\pm f_k\vert}\, .
\end{equation}
The Zak phase is therefore simply
\begin{equation}
\Phi=\frac{i}{2}\sum_{\tau=\pm 1}\int_{-\pi}^\pi dk\, e^{i\varphi_k^{(\tau)}}\partial_ke^{-i\varphi_k^{(\tau)}}=\frac{1}{2}\left[\varphi_k^{(+)}+\varphi_k^{(-)}\right]_{-\pi}^{\pi}\, .
\end{equation}
The $2\pi$-periodicity of $\varepsilon_k$ and $f_k$ implies that $\varphi_{-\pi}^{(\tau)}$ and $\varphi_{+\pi}^{(\tau)}$ differ by a multiple of $2\pi$. Therefore the only possible values of the Zak phase are multiples of $\pi$.

\begin{figure}[htp!]
	\centering
	\includegraphics[width=.5\textwidth]{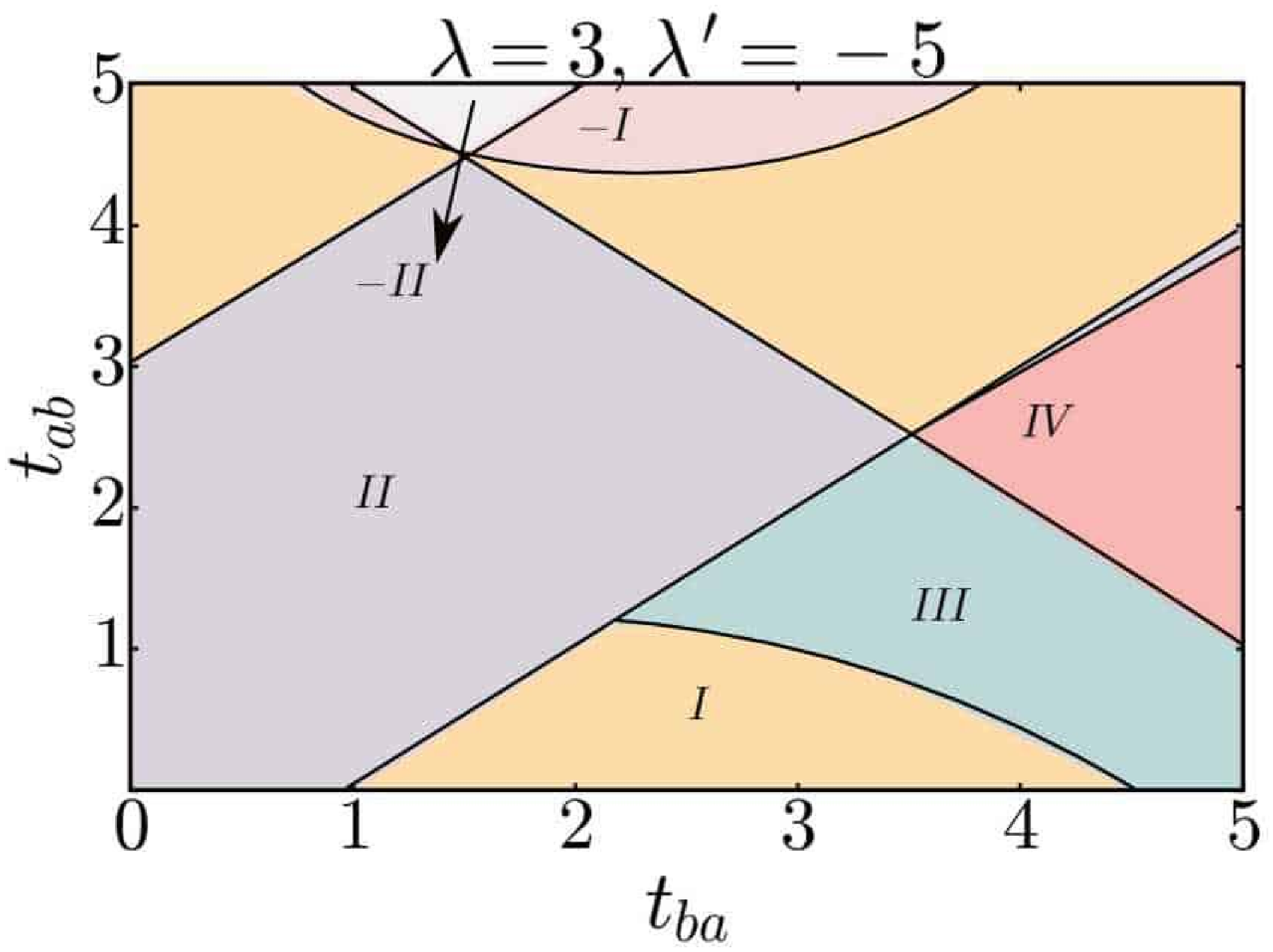}
	\includegraphics[width=.5\textwidth]{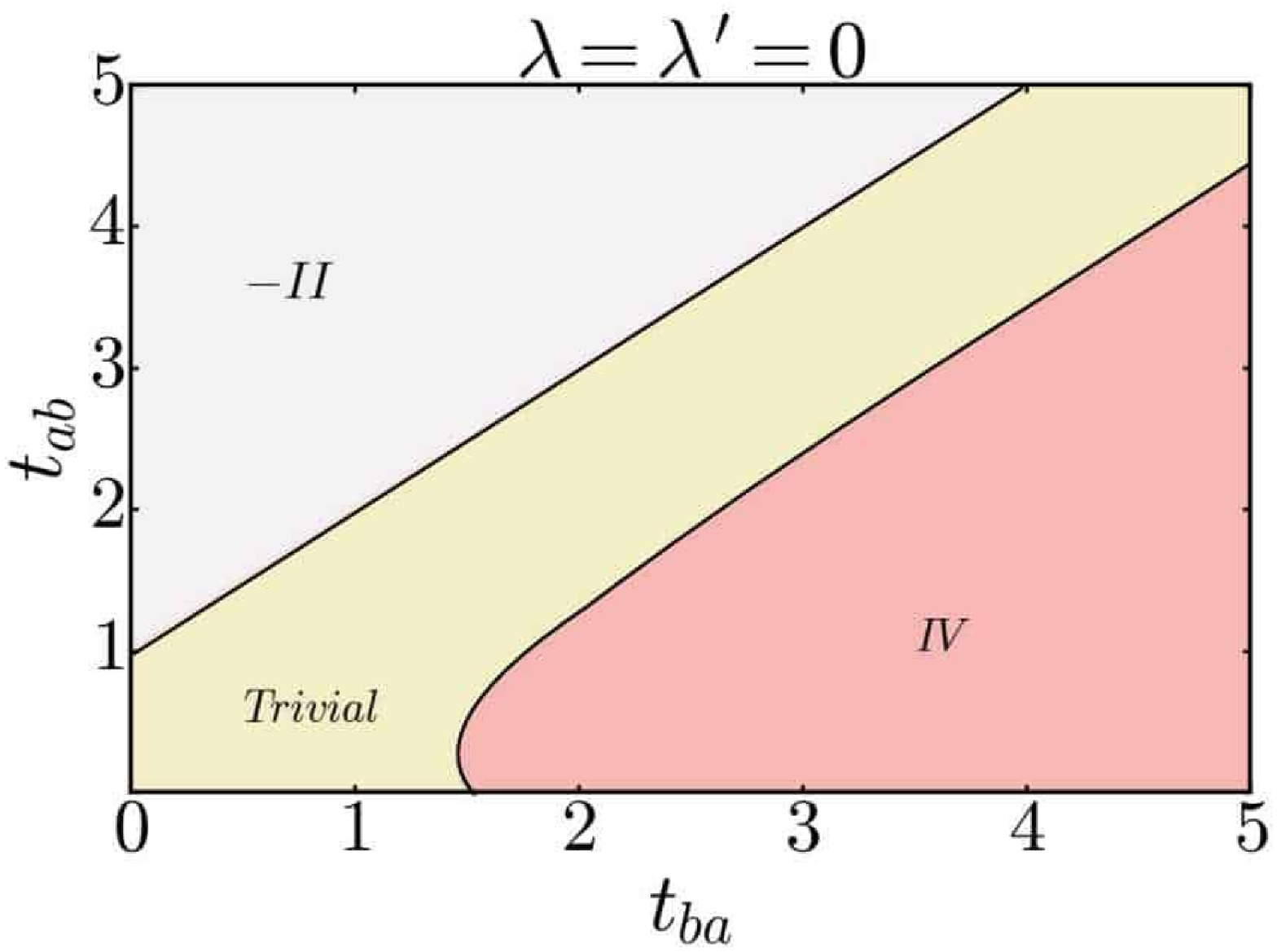}
	\caption{(Color online) Ground-state phase diagram of model (\ref{Eq:ODDHamiltonian}), for $\delta=0.5$ and either finite (top) or vanishing SOC (bottom).} 
	\label{Fig:3rd-phase-diag}
\end{figure}

\begin{figure}[h!]
	\centering
	\includegraphics[width=0.48\textwidth]{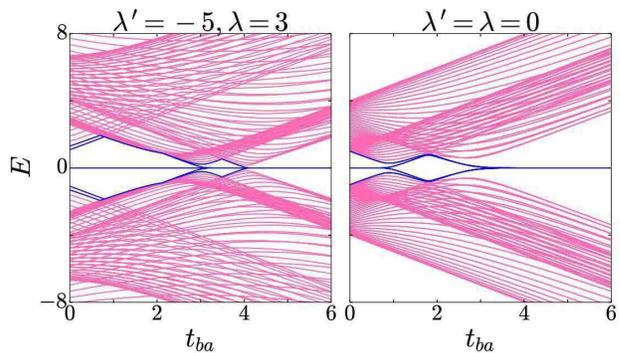}
	\caption{(Color online) Band structure of model (\ref{Eq:ODDHamiltonian}) for a finite chain of 100 sites (open boundary conditions), and for 
	$\delta=0.5$, $t_{ab}=2$. The figure on the left refers to a horizontal line in the upper phase diagram of Fig. \ref{Fig:3rd-phase-diag}, the one to the right corresponds to the lower phase diagram.}
		\label{Fig:Edge3rdN}
\end{figure}

We find consistently that for a Zak phase $\Phi=\nu\pi$ there are $\vert\nu\vert$ zero-energy edge modes (bulk-boundary correspondence). The ``total winding number'' $\nu$ is well suited for labelling the various phases. In fact, the phase diagram in the $t_{ba}-t_{ab}$ plane, shown in Fig. \ref{Fig:3rd-phase-diag}, is a complex patchwork of regions having different values of $\nu$ (up to $\nu=IV$). The patches are separated from each other either by straight lines (for a gap closing at $k=0$ or $k=\pi$) or by curved boundaries (on which the gap closes at $k\neq 0,\pi$). 
The transitions between two topological phases is always continuous because of the smooth evolution of the band structure as a function of parameters, as exemplified in Fig. \ref{Fig:Edge3rdN}.


\begin{figure}[htp!]
	\centering
	\begin{tabular}{l}
	\includegraphics[width=.4\textwidth]{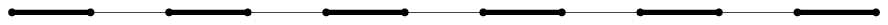}\\ \\
	\includegraphics[width=.4\textwidth]{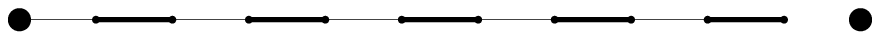}\\ \\
	\includegraphics[width=.4\textwidth]{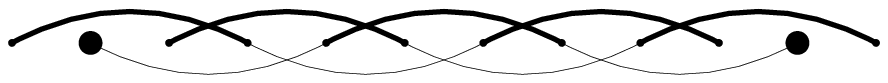}\\ \\
	\includegraphics[width=.4\textwidth]{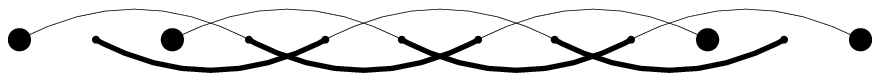}
	\end{tabular}
	\caption{Bond picture and edge states for a chain with 12 sites. The first two cases represent the bond alternation of the SSH model 
	($ t_{ab}$, $ t_{ba}\ll 1$), where the upper (lower) figure has strong (weak) bonds at the edges. The other two cases are for dominant 
	third-neighbor hopping ($t_{ab}$, $ t_{ba}\gg 1$) with $t_{ab}>t_{ba}$ (upper figure) and $t_{ab}<t_{ba}$ (lower figure).}
	\label{Fig:bonding}
\end{figure}

Fig. \ref{Fig:3rd-phase-diag} can be qualitatively understood in terms of the ``chemical structures'' shown in Fig. \ref{Fig:bonding}. In the absence of SOC and for weak third-neighbor hopping we obtain bond alternation without edge states for $\delta>0$ (as assumed here), but two edge modes for 
$\delta<0$. If third-neighbor hopping dominates, the bonding picture exhibits two edge modes if $t_{ab}$ is the dominant term, but four if $t_{ba}$ is 
larger, in agreement with Fig. \ref{Fig:3rd-phase-diag}. 

In Eq. (\ref{Eq:vectors}) we have made a particular choice for the phases of the vector components. Instead of choosing the first and third components as real, we could have done so for the second and fourth components. This would give a phase factor $e^{i\varphi_k^{(\pm)}}$ for the other components, and therefore $\Phi$ would change sign. The phase diagram of course would again be well repoduced in terms of winding numbers and the bulk-boundary correspondence would not be affected. However, if we choose an entirely complex representation, with phase factors $e^{i\varphi_k^{(\pm)}/2}$ for the first and third
and $e^{-i\varphi_k^{(\pm)}/2}$ for the second and fourth components, the Zak phase vanishes and does not give any information about the phase diagram. Therefore the "right" gauge choice seems to be a mixed representation. We will see later that the entirely complex representation is the natural choice in cases where the Zak phase is not quantized, but can be related to some local observable.
%%%%%%%%%%%%%%
\subsection{Entanglement}\label{sec:entanglement}
%%%%%%%%%%%%%%%%%%%%%%%%%%%%%%%%%%%%%%%%%%%%%%%%%%%%%%%%%%%%%%%%%%%%
Entanglement recently became an important tool for characterizing quantum many-body states \cite{Laflorencie_16}. To define this concept quantitatively, one divides a system into two parts, a ``block'' and an ``environment'', and calculates the reduced density matrix for the degrees of freedom of the block by tracing out those of the environment. In the context of independent (spinless) fermions a remarkably simple expression has been derived for the reduced density matrix $\rho_N$ for a block of $N$ sites in terms of the so-called correlation matrix $G_N$ (to be specified below) \cite{PhysRevB.64.064412,PhysRevB.69.075111,Peschel_2009},
\begin{equation}
\rho _N = \det\left(1-G_N\right)\exp[-\sum_{\ell}\phi_\ell \dg
c_\ell c_\ell] = \frac 1Z\exp[-H_{\mbox{\scriptsize ent}}].\label{Eq:rhoN-2}
\end{equation}
Here $H_{\mbox{\scriptsize ent}}=\sum_{\ell}\phi_\ell \dg c_\ell c_\ell$ is the
so-called entanglement Hamiltonian, and the single-particle {\it entanglement spectrum} (ES) is given by
\begin{equation}
\phi_{\ell}=-\left(\log G_N\left(1-G_N\right)^{-1}\right)_{\ell\ell}
= -\log \frac{\eta_\ell}{1- \eta_\ell}, \label{Eq:phi}
\end{equation}
in terms of the eigenvalues $\eta_{\ell}$ of $G_N$. 

In our case the spin degrees of freedom have to be taken into account explicitly because of the SOC. The correlation matrix $G_N$ therefore has $2N\times 2N$ matrix elements. At half filling these are given by the single-particle correlation functions
\begin{equation}
\nonumber \la \dg a_{i\s}a_{j\s'}\ra= \la \dg b_{i\s}b_{j\s'}\ra
=\frac{1}{2} \delta_{ij}\delta_{\s\s'},
\end{equation}
\begin{eqnarray}
\nonumber\la \dg a_{i\s}b_{j\s}^{\phantom{}}\ra&=&\frac{-1}{8\pi}\int_{-\pi}^{\pi}dke^{-ikd}\left(\frac{1}{E_k^-}+\frac{1}{E_k^+}\right),\\
\nonumber\la \dg a_{i\s}b_{j-\s}^{\phantom{}}\ra&=&\frac{1}{8\pi}\int_{-\pi}^{\pi}dke^{-ikd}\left(\frac{1}{E_k^-}-\frac{1}{E_k^+}\right),\\
\nonumber\la \dg b_{i\s}a_{j\s}^{\phantom{}}\ra&=&\frac{-1}{8\pi}\int_{-\pi}^{\pi}dke^{-ikd}\left(E_k^-+E_k^+\right),\\
\la \dg b_{i\s}a_{j-\s}^{\phantom{}}\ra
&=&\frac{1}{8\pi}\int_{-\pi}^{\pi}dke^{-ikd}\left(E_k^--E_k^+\right),\label{Eq:corr-function-SSH+SO}
\end{eqnarray}
where $E_k^\pm=\vert\ve_k\pm f_k\vert$, and $d=|i-j|$.
By computing these integrals we can obtain the ES of the Hamiltonian (\ref{Eq:ODDHamiltonian}) from Eq. (\ref{Eq:phi}). 

\begin{figure}[h!]
	\centering
	\includegraphics[width=.48\textwidth]{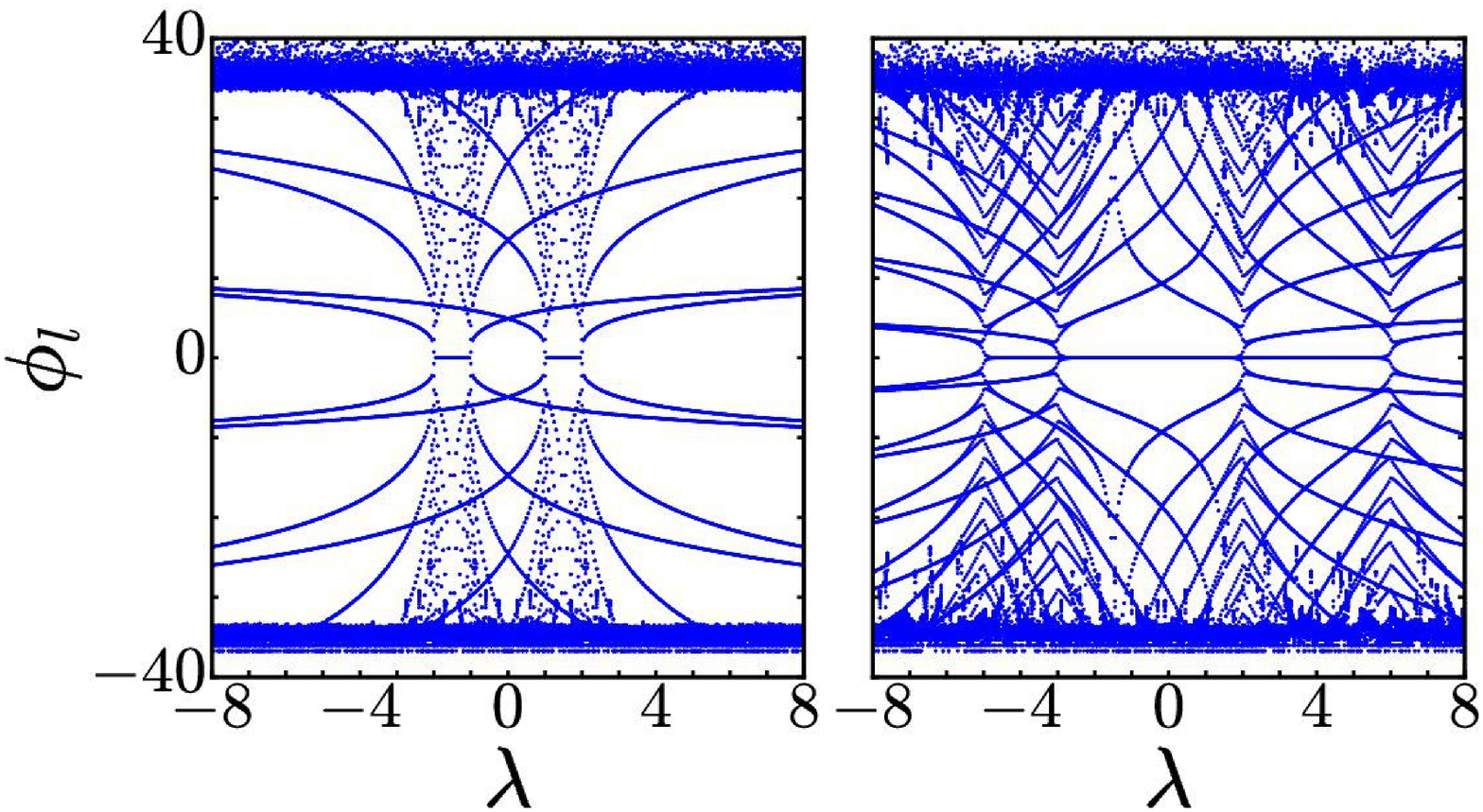}
	\includegraphics[width=.48\textwidth]{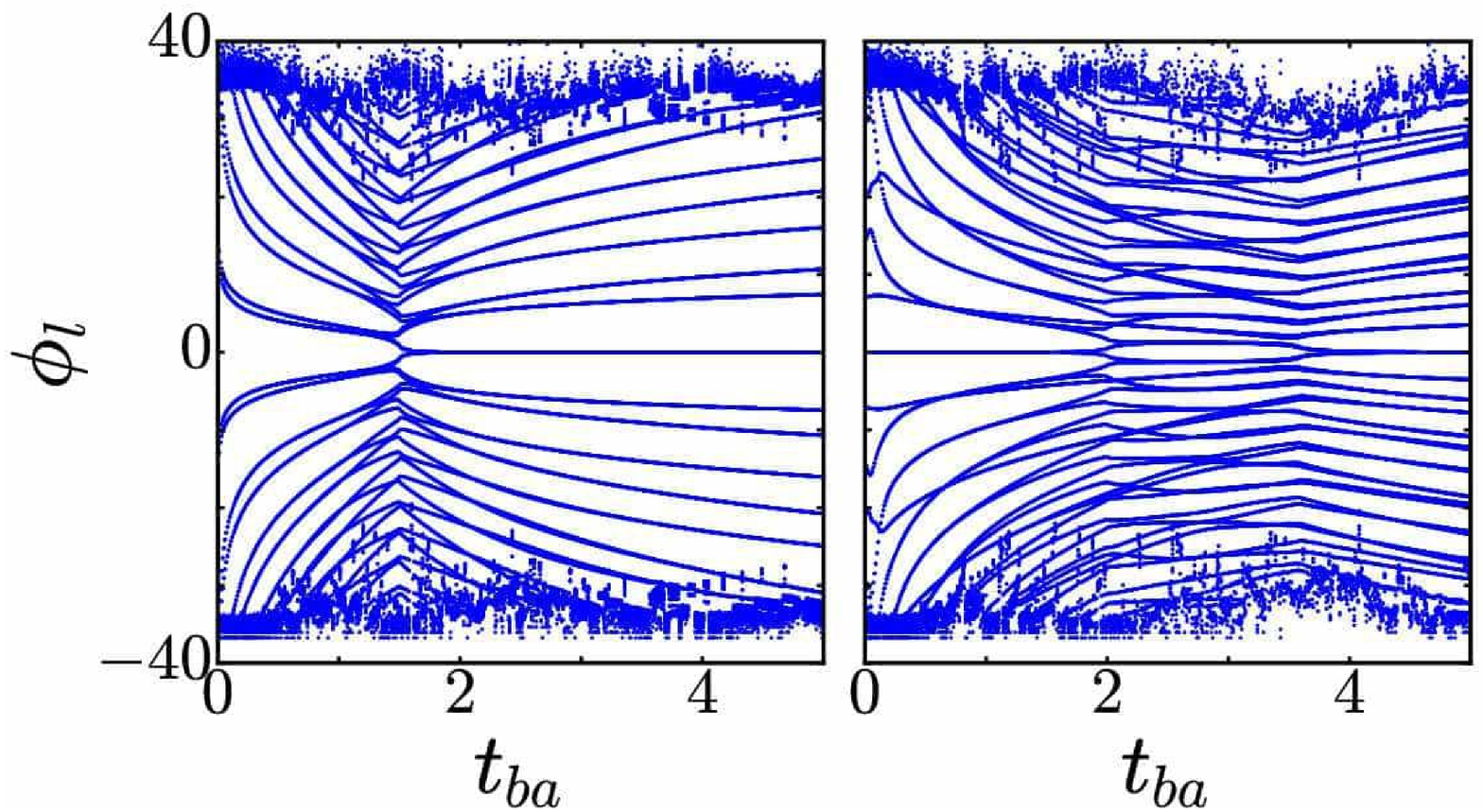}
\caption{(Color online)
 ES of model (\ref{Eq:ODDHamiltonian}), computed for a finite block of $60$ sites. Top row: No third-neighbor hopping, $\delta =0.5$, $\lambda'=0$ (left), $\lambda'=4$ (right). Bottom row: Vanishing SOC, but finite third-neighbor hopping, $t_{ab}=0.5$ with a critical point at $t_{ba}=1.51$ (left), $t_{ab}=3$ with critical points at $t_{ba}=2$ and $3.59$ (right).}
	\label{Fig:ent-spect-SSH+SO}
\end{figure}

Fig. \ref{Fig:ent-spect-SSH+SO} shows the ES of the extended SSH model with and without third-neighbor hopping.
Different phases correspond to different patterns of the ES, and critical points occur where these patterns change drastically.  Particle-hole symmetry is also manifest. An interesting observation on the zero modes is worth mentioning. They appear at the same parameter values in the ES as in the band structure of the full Hamiltonian, and the numbers of zero modes are identical. 
%%%%%%%%%%%%%%%%%%%%%%%%%%%%%%%%%%%%%%%%%%%%%%%%%%%%%%%%%%%%%%%%%%%%%%%%%%%%
%%%%%%%%%%%%%%%%%%%%%%%%%%%%%%%%%%%%%%%%%%%%%%%%%%%%%%%%%%%%%%%%%%%%%%%%%%%%%

The ES can give a qualitative picture of the various phases. In order to gain more accurate information, for instance about the critical points, one has to consider quantities which are more specific. One of them is the {\it entanglement gap}, defined as
the difference between the lowest positive level and the highest
negative level as\cite{PhysRevB.84.195103}
\begin{equation}
\Delta _E = \left| \phi_\ell^+ -\phi_\ell^- \right|.
\end{equation}
\begin{figure}
    \centering
    \includegraphics[width=.48\textwidth]{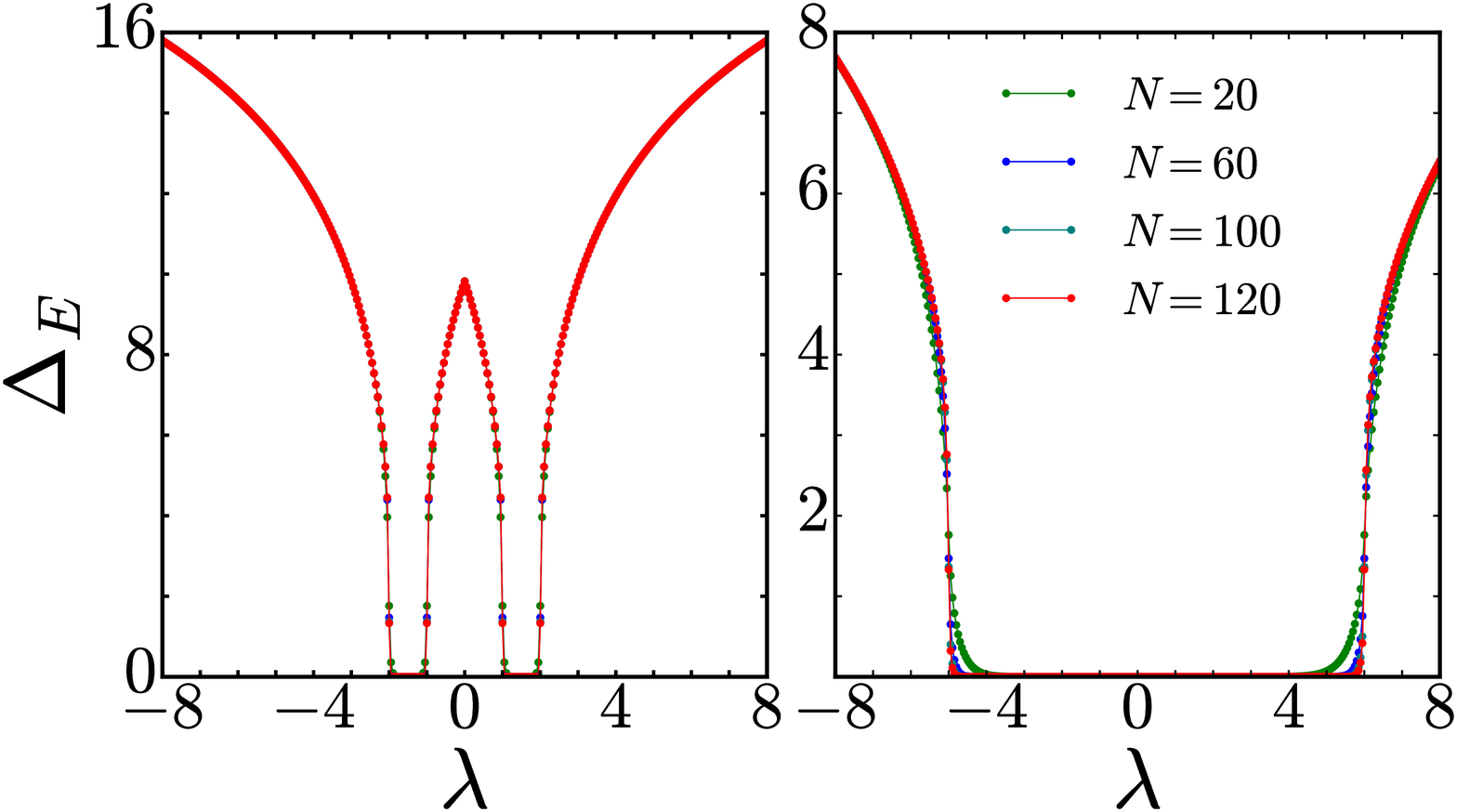}
    \includegraphics[width=.48\textwidth]{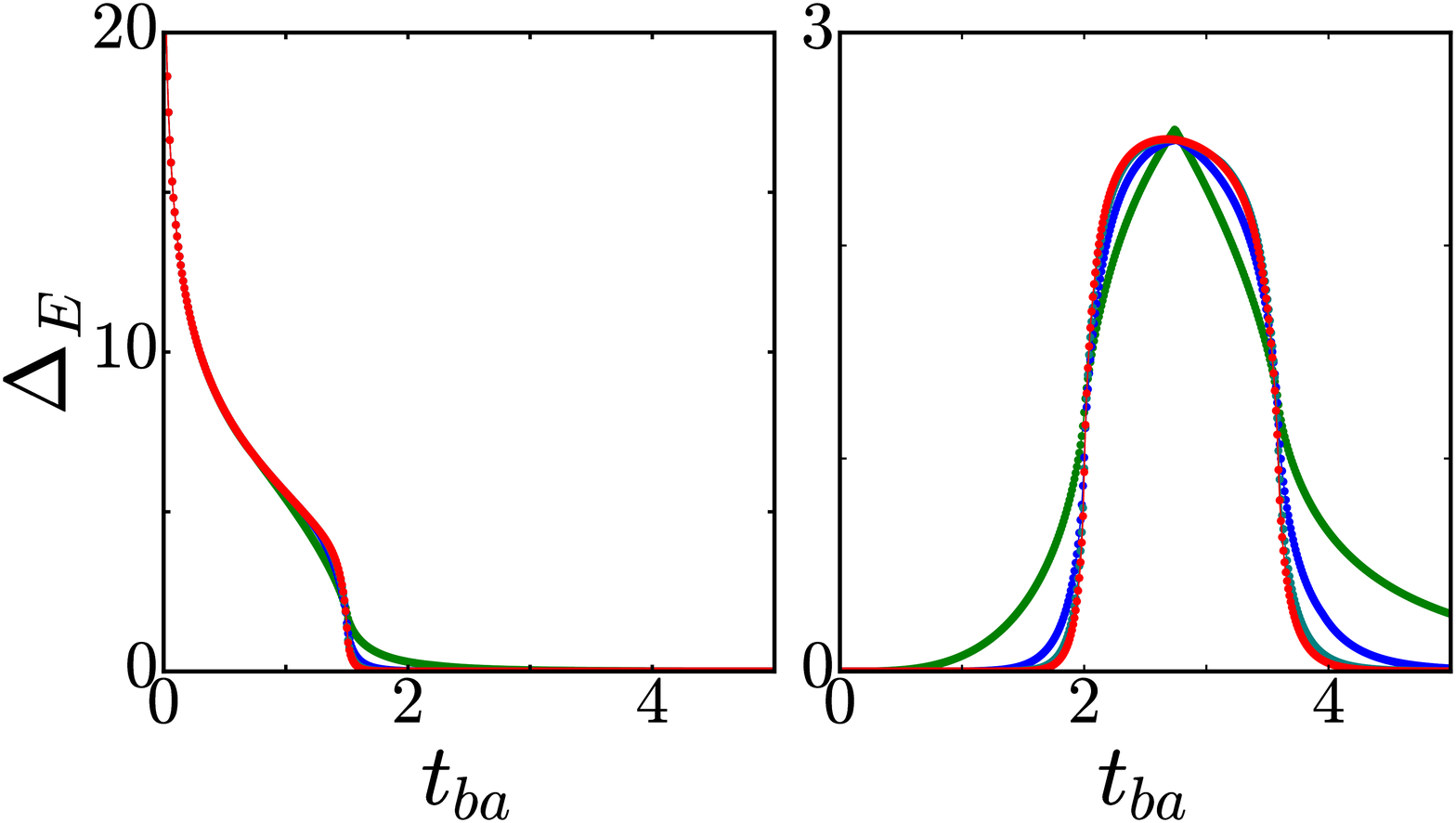}
    \caption{(Color online) Entanglement gap for $\delta =0.5$ and different subsystem sizes. Top: No third-neighbor hopping, $\lambda'=0$ $(4)$ in the left (right) plot. Bottom: Finite third-neighbor hopping, $\lambda=\lambda'=0$ for $t_{ab}=0.5$ $(3)$, in the left (right) plot.}
    \label{Fig:ent-gap-SSH+SO}
\end{figure}
%%%%%%%%%%%%%%%%%%%%%%%%%%%%%%%%%%%%%%%%%%%%%%%%%%%%%%%%%%%%%%%%%%%%%%%%%%%%%%
Fig. \ref{Fig:ent-gap-SSH+SO} shows the entanglement gap as a function of both SOC and third-neighbor hopping. The gap is finite in topologically trivial phases (with no zero modes) and vanishes in non-trivial phases. To see a sharp onset of a finite gap, one has to choose large subsystem sizes.

\begin{figure}
    \centering
    \includegraphics[width=0.5\textwidth]{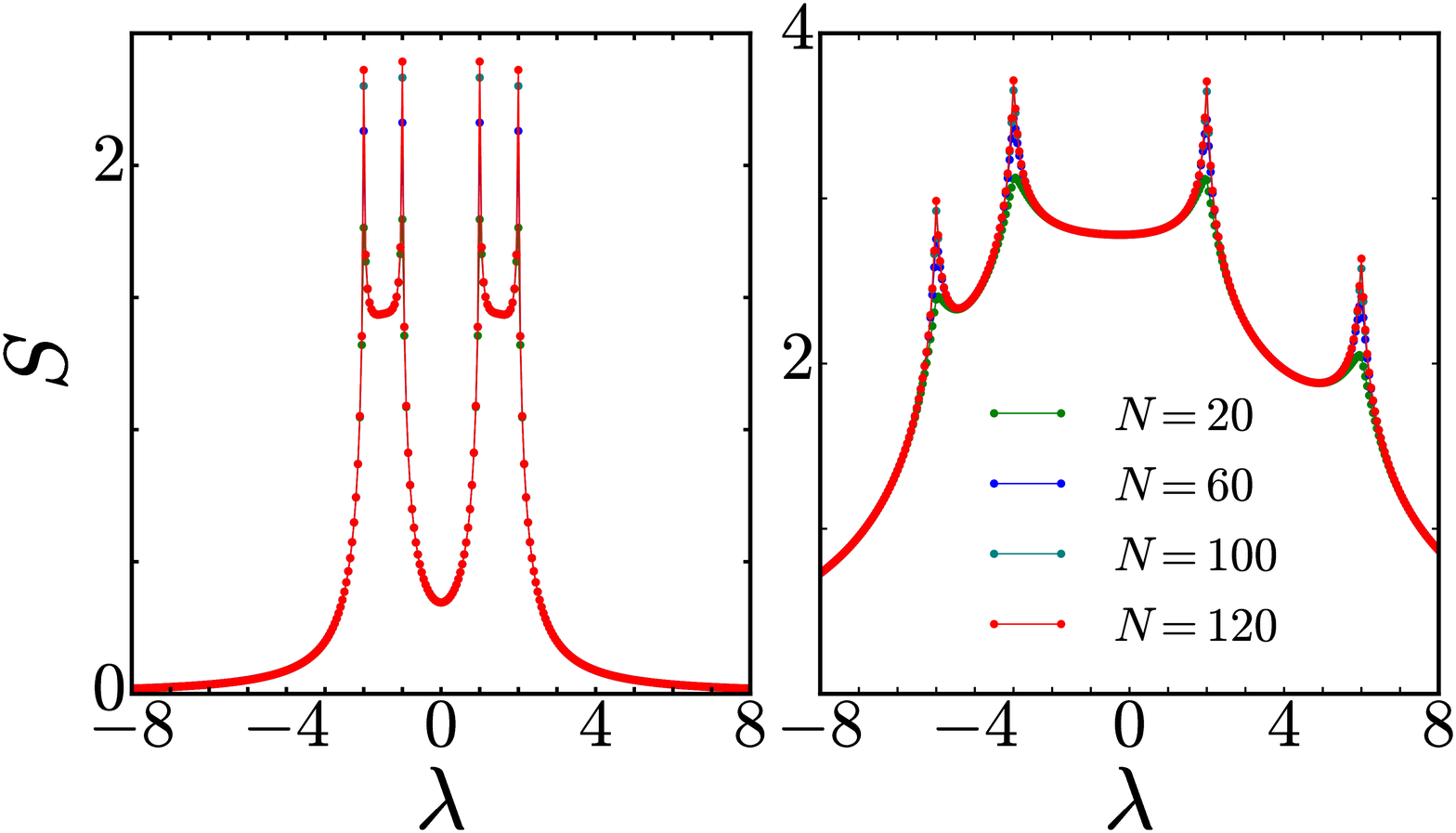}
    \includegraphics[width=0.5\textwidth]{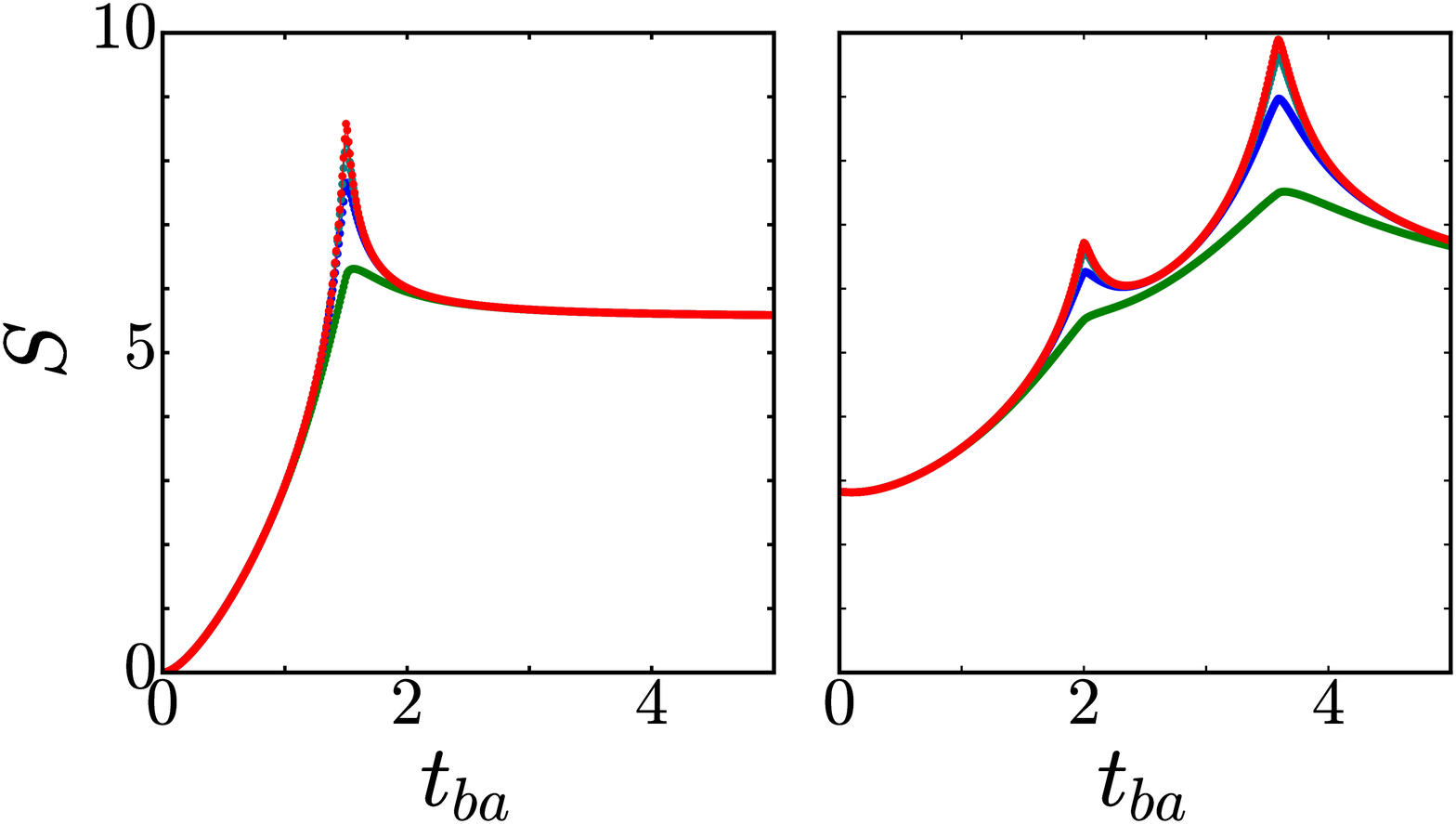}
    \caption{(Color online) 
    	Entanglement entropy for $\delta t=0.5$ and different subsystem sizes. Top: No third-neighbor hopping, $\lambda'=0$ $(4)$ in the left (right) plot. Bottom: Finite third-neighbor hopping, $\lambda=\lambda'=0$, $t_{ab}=0.5$ $(3)$, in the left (right) plot.}
    \label{Fig:EE-SSH+SO}
\end{figure}

Another useful quantity is the {\it entanglement entropy} $S$, which can be defined in terms of the eigenvalues of the correlation matrix $G_N$,
\begin{equation}
S= -\sum_\ell\eta_\ell\log\eta_\ell -\sum_\ell\left(1-\eta_\ell\right)\log\left(1-\eta_\ell\right).
\end{equation}
Eigenvalues close to 0 or 1 have little weight, and $S$ is dominated by a small interval around $\eta = \frac 12$. 
Away from critical points, the spectrum of the correlation matrix is gapped, eigenvalues are close to either 0 or 1, and $S$ is small. 
Close to a critical point, a significant part of the eigenvalues is clustered around 1/2, and $S$ diverges at criticality for $N\rightarrow\infty$ (see Fig. \ref{Fig:EE-SSH+SO}). This will be further discussed in the next subsection. 
%%%%%%%%%%%%%%%%%%%%%%%%%%%%%%%%%%%%%%%%%%%%%%%%%%%%%%%%%%%%%%%%%%%%%%%%%%%%%%
%%%%%%%%%%%%%%%%%%%%%%%%%%%%%%%%%%%%%%%%%%%%%%%%%%%%%%%%%%%%%%%%%%%%%%%%%%%%%%%

By analyzing the
eigenvalues of the correlation matrix $G_N$, we realize that in the topologically trivial
phase all eigenvalues of $G_N$ are localized in narrow intervals
around zero and one, while in SPT-II a non-negligible portion of
eigenvalues is clustered around $0.5$. By contrast, in SPT-I the distribution of eigenvalues around 0 and 1 is wider. So it is clear that 
$S_{II}>S_{I}>S_{\mbox{\scriptsize trivial}}$
 (see also Fig. \ref{Fig:EE-SSH+SO}).
On the other hand, $\Phi_{II}> \Phi_{I}> \Phi_{\mbox{\scriptsize trivial}}$. These two relations indicate that 
the larger the Berry phase $\Phi$, the larger the entanglement entropy $S$, a possible consequence of the bulk-edge correspondence \cite{PhysRevB.73.245115}.

%%%%%%%%%%%%%%%%%%%%%%%%%%%%%%%%%%%%%%%%%%%%%%%%%%%%%%%%%%%%%%%%%%%%%%%%%%%%%%%%%%%%%%
%%%%%%%%%%%%%%%
\subsection{Critical behavior}\label{sec:critical}
%%%%%%%%%%%%%%%
The lack of an order parameter at a topological phase transition requires alternative quantities to be considered. The fidelity susceptibility has been widely used \cite{Albuquerque_10, Luo_14, 1742-5468-2014-10-P10032, Koenig_16}; it represents energy fluctuations in the ground state and thus is something like a quantum specific heat. Another important quantity is the correlation length $\xi$, which has been extracted from the penetration of edge states into the bulk \cite{Rufo_19}. A different scheme for calculating $\xi$ has been proposed in terms of the Berry connection $(\langle u_k| i \partial _k u_k\rangle)$ in one dimension and the Berry curvature in two dimensions; their Fourier transforms can be interpreted as real-space correlation functions involving the electric polarization and orbital currents in one and two dimensions, respectively \cite{Chen_17, Chen_19}.  

The distinction between conventional quantum phase transitions (with symmetry breaking) and topological transitions is not always clear-cut. 
As discussed in Section \ref{sec:intro}, the SSH Hamiltonian (\ref{Eq:ham_SSH}) exhibits a topological transition at $\delta=0$ from a topologically trivial to a topologically non-trivial phase. At the same time, this transition can also be understood in terms of symmetry breaking. In fact, translational symmetry is broken by $\delta\neq 0$, leading to bond alternation with different sequences for different signs of $\delta$.
For the generalized SSH Hamiltonian (\ref{Eq:ODDHamiltonian}), the phase diagram exhibits a variety of phases, characterized by different winding numbers, as illustrated in Fig. \ref{Fig:3rd-phase-diag}. These phases in general do not differ with respect to symmetry and therefore the transitions are of genuine topological nature. Fortunately, in the present case it is very simple to determine the critical behavior, which depends only on the single-particle spectrum close to the $k$-point where the gap closes at criticality \cite{Rufo_19}.

Along the critical lines of Fig.\ref{Fig:3rd-phase-diag}, the low-energy single-particle spectrum has in general a linear dispersion and the critical behavior should therefore be that of a conformal field theory with central charge $c=1$ \cite{Henkel_99}. This fundamental quantity, which is ``(almost) sufficient to characterize a critical model'' \cite{Itzykson_89}, can be calculated from the entanglement entropy $S$. 
It is known that at criticality the entanglement entropy of a block of size $N$ scales as 
$S(N)=\frac{c}{3}\log N+S_0$, where $S_0$ is a non-universal constant \cite{PhysRevLett.90.227902}. 
We have calculated $S$ as a function of block size $N$ in the absence of third-neighbor hopping, $t_{ab}=t_{ba}=0$, for various critical points. Along the critical lines we consistently find $c=1$, as expected. But what happens at the crossing of two such lines, for instance for $\delta=0.5, \lambda=0.5, \lambda'=-1.5$, where four phases meet, two topological phases with $\nu=I$, one with $\nu=II$ and a trivial phase? The low-energy spectrum has two Dirac cones, one at $k=0$, the other at $k=\pi$. Therefore we get two fields and a central charge $c=2$. This is indeed what we also obtain from the entanglement entropy.  

%%%%%%%%%%%%%%%%%%%%%%%%%%%%%%%%%%%%%%%%%%%%%%%
\section{Effects of broken charge-conjugation symmetry}\label{sec:SO-SSH+Even}

So far we have limited ourselves to systems with hopping between different sublattices (``odd hopping'') and therefore with charge-conjugation (particle-hole) symmetry. In this section we investigate the effect of next-nearest-neighbor hopping, which breaks this symmetry. The Hamiltonian is
\begin{equation}\label{Eq:Ham2}
H=H_{\mbox{\scriptsize SSH}}+H_{\mbox{\scriptsize SO}}+H_{\mbox{\scriptsize NNN}}+H_{\mbox{\scriptsize NNNN}}\, ,
\end{equation}
where $H_{\mbox{\scriptsize SSH}}$ and $H_{\mbox{\scriptsize NNNN}}$ are defined by Eqs. (\ref{Eq:ham_SSH}) and (\ref{Eq:ham_NNNN}), respectively, $H_{\mbox{\scriptsize SO}}$ is the spin-orbit coupling (\ref{Eq:ham_SOC})
and $H_{\mbox{\scriptsize NNN}}$ is the additional hopping term, connecting sites of the same sublattice,
\begin{equation}\label{Eq:ham_NNN}
H_{\mbox{\scriptsize NNN}}=\sum_{i\sigma}\left(t_{a}\, a_{i\s}^{\dagger}a_{i+1\s}^{\phantom{}}+  t_{b}\, b_{i\s}^{\dagger}b_{i+1\s}^{\phantom{}} + \mbox{h.c.}\right).
\end{equation}

\begin{figure}[ht]
	\centering
    \includegraphics[width=0.5\textwidth]{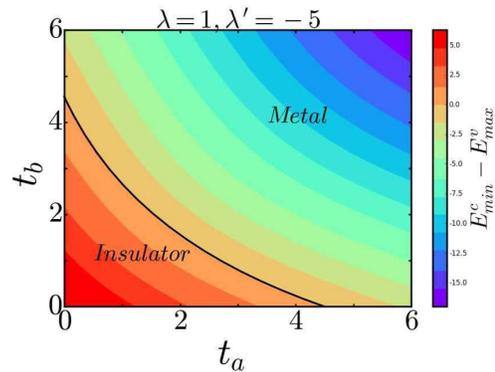}
	\caption{ (Color online) Band gap as a function of the hopping amplitudes $t_a$ and $t_b$ for $\delta=0.5$, $\lambda=1$, $\lambda'=-5$ and 
	$t_{ab}=t_{ba}=0$. The (black) solid line marks the boundary where the band gap vanishes. This is a transition line between insulating and metallic phases.}
	\label{Fig:Metal_Ins.tran}
\end{figure}

The addition of $H_{\mbox{\scriptsize NNN}}$ destroys the symmetry between positive- and negative-energy levels, or between conduction and valence bands. As shown in Fig. \ref{Fig:Metal_Ins.tran}, the band gap, i.e. the spacing between the conduction-band minimum and the valence-band maximum, decreases as $t_a$ and $t_b$ increase, and it closes on a critical line, which signals a transition from an insulating to a (semi-) metallic phase. The interesting region for our study is the insulating phase where the spectrum is gapped at the Fermi energy. Therefore the hopping amplitudes between next-nearest neighbors should not be too large.

In the presence of second-neighbor hopping, time-reversal symmetry $T$ is preserved, but this is not the case for charge conjugation $C$ nor for the chiral symmetry $S$.
Therefore the Hamiltonian (\ref{Eq:Ham2}) belongs to the AI symmetry class. According to the periodic table of topological insulators/superconductors\cite{PhysRevB.55.1142,PhysRevB.78.195424,PhysRevB.78.195125}, this class is topologically trivial. Nevertheless, some properties resemble features found in topological phases, especially if the hopping parameters $t_a$ and $t_b$ are chosen in such a way that the system keeps a spatial symmetry. We consider the cases $t_a=t_b$, where parity $P$ is conserved and $t_a=-t_b$, where the product $CP$ is conserved, while individually both $C$ and $P$ are broken.
%%%%%%%%%%%%%%%%%%%%%%%
\subsection{Conserved parity}\label{sec:parity}
%%%%%%%%%%%%%%%%%%%%%%%

\begin{figure}[h!]
	\centering
	\includegraphics[width=0.4\textwidth]{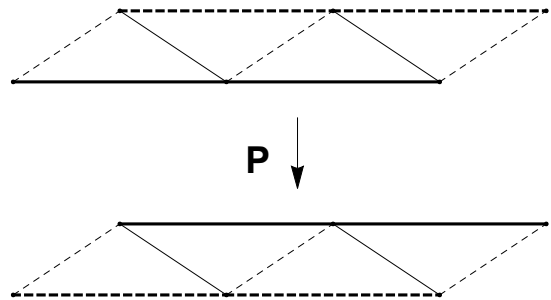}
	\caption{Effect of the parity operation on a chain with both nearest-neighbor (thin lines) and next-nearest-neighbor bonds (thick lines). The initial configuration is restored if the second-neighbor hopping amplitudes are equal in the two sublattices. Third-neighbor hopping (not shown) does not change the argument.}
	\label{Fig:nnbonds}
\end{figure}

Fig. \ref{Fig:nnbonds} illustrates the connectivity induced by nearest- and next-nearest-neighbor hopping, and it shows how a given pattern changes due to a parity operation, where $b$-sites at $i$ are exchanged with $a$-sites at $L+1-i,\, i=1,...,L$. Clearly the pattern is invariant if $t_a=t_b$.
For the Bloch Hamiltonian $H_k$, which is again a $4\times 4$ matrix as in Eq. (\ref{Eq:Bloch}), with additional elements $2t_a\cos k$ and $2t_b\cos k$ on the diagonal, parity corresponds to the operator $i\kappa\sigma_0\otimes \sigma_x$. One finds $PH_kP^{-1}=H_k$, provided that $t_a=t_b$. 

\begin{figure}[h!]
	\centering
	\includegraphics[width=0.5\textwidth]{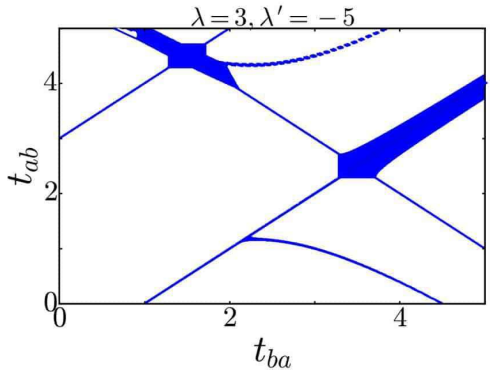}
	\caption{(Color online) Phase diagram for the same parameters as in Fig. \ref{Fig:3rd-phase-diag} plus NNN hopping with $t_a=t_b=0.1$. Metallic regions are shaded.}
	\label{Fig:pd_NNN}
\end{figure}

The eigenvalues of the Bloch Hamiltonian are $2t_a\cos k+\vert\varepsilon_k\pm f_k\vert$ (``conduction bands'') and $2t_a\cos k-\vert\varepsilon_k\pm f_k\vert$  (``valence bands''). The phase diagram is determined by calculating the difference between conduction band minima and valence band maxima. For finite $t_a$ this difference not only can vanish (``gap closing'') but it can be negative, as already illustrated in Fig. \ref{Fig:Metal_Ins.tran}. Even for weak NNN hopping this can happen in the vicinity of transition lines, which may broaden into metallic patches. In fact, some of the topological transitions found for $t_a=0$ are split and replaced by pairs of metal-insulator transitions, as shown in Fig. \ref{Fig:pd_NNN}.

Interestingly, in the gapped regions of the phase diagram, where valence and conduction bands are separated, the Zak phase remains quantized even in the presence of NNN hopping. Indeed, for the lowest two eigenvalues $2t_a\cos k-\vert\varepsilon_k\pm f_k\vert$ the eigenstates $\vert u_k\rangle$ satisfy exactly the same equations as for $t_a=0$, and therefore the Zak phase is not affected.

\begin{figure}[h!]
	\centering
	\includegraphics[width=0.5\textwidth]{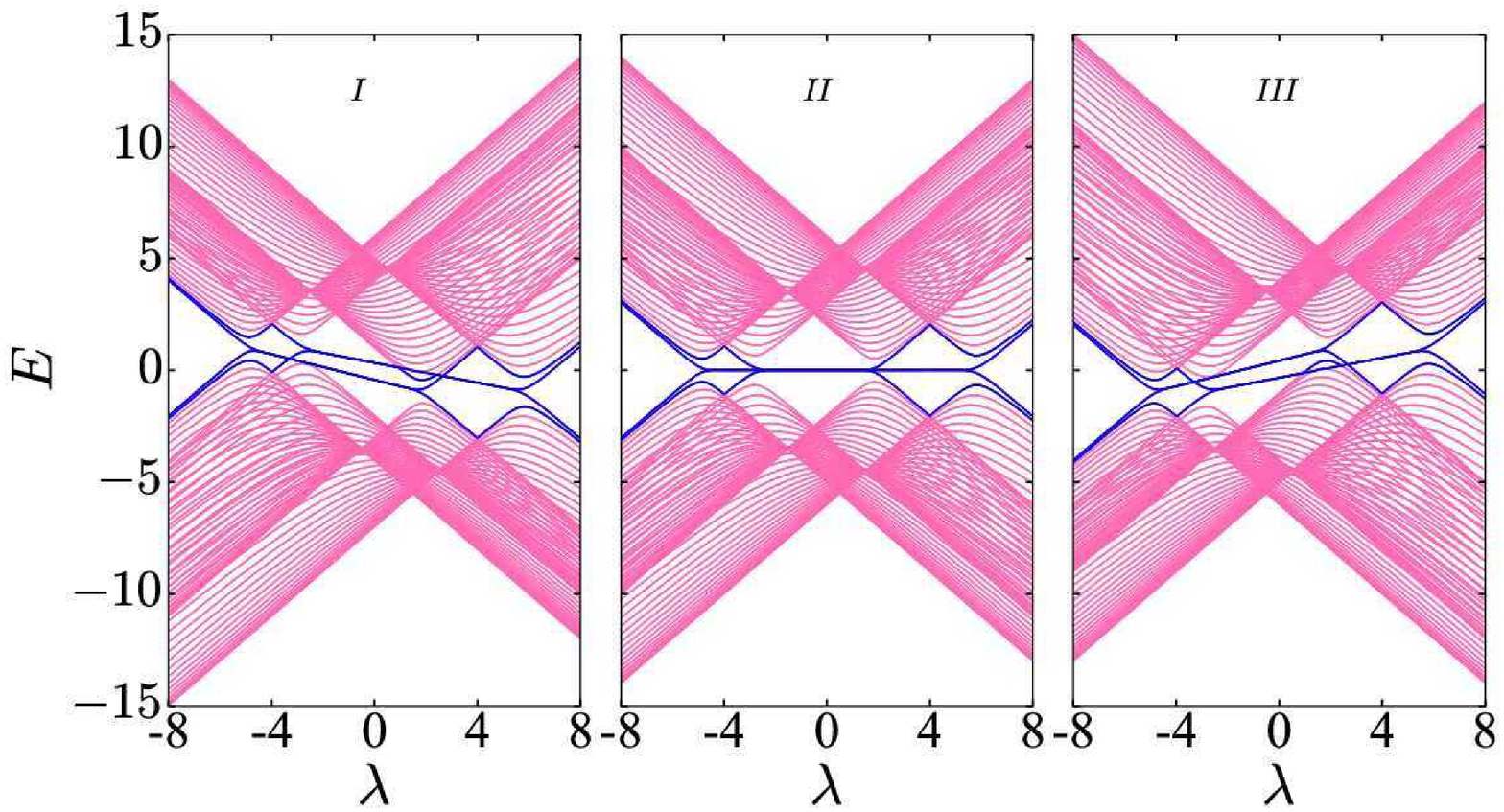}
	\includegraphics[width=0.5\textwidth]{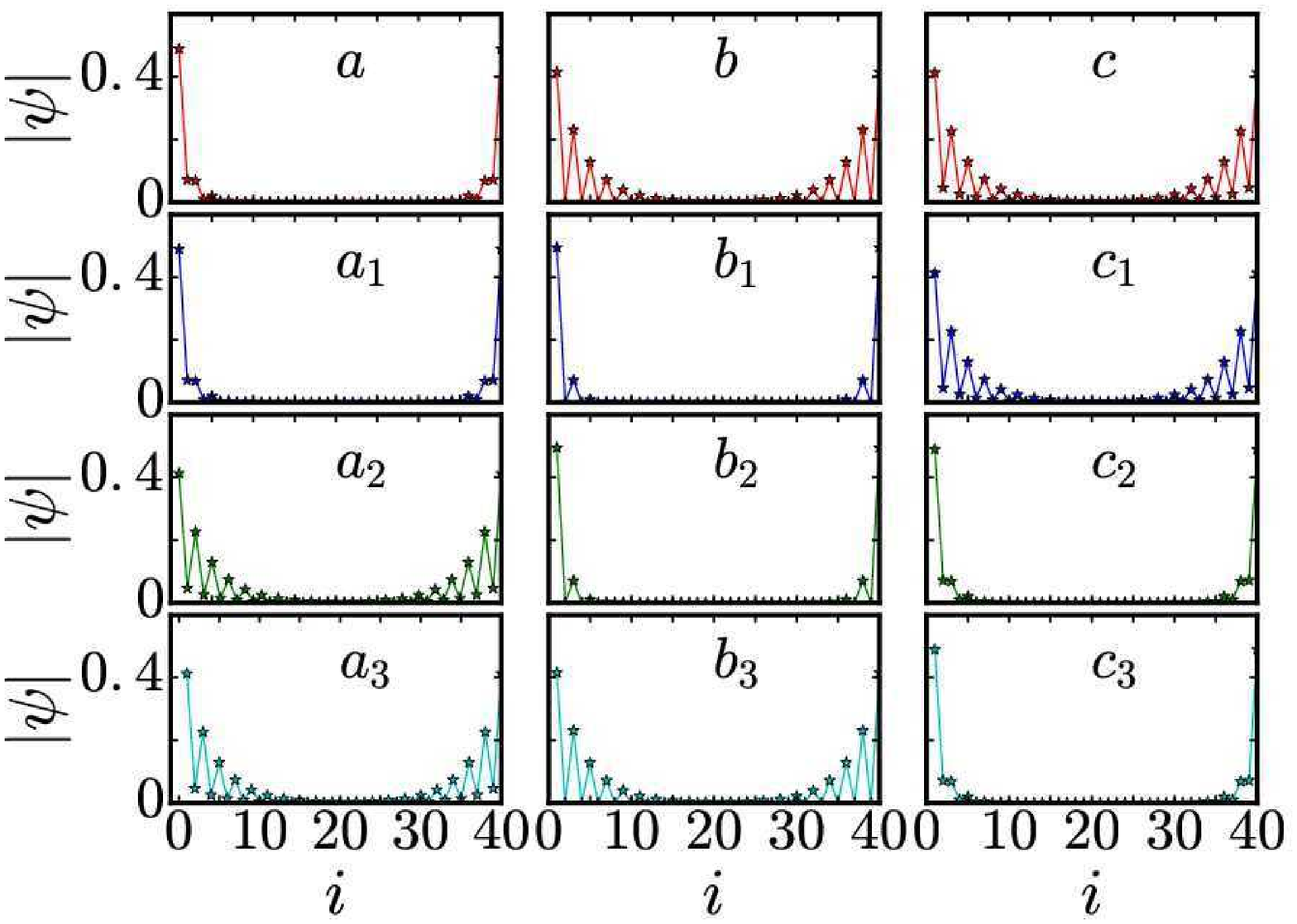}
	\caption{(Color online) Top: Energy spectrum of a chain with $L=60$ sites versus $\lambda$, for $\lambda'=4$ and different NNN hopping strengths: (I) $t_a=t_b=-0.5$, (II) $t_a=t_b=0$, and (III) $t_a=t_b=0.5$. Bottom: Probability amplitude for edge states along a chain of size $L=40$, for $\lambda=-1$ and $\lambda'=4$. The energies of the edge states are: $E=0.55$ and $-0.142$, respectively, for the upper and lower plots on the left, $E=0.142$ and $-0.55$, respectively, for the upper and lower plots on the right, and $E=0$ for the plots in the middle.}
	\label{Fig:Edgestateplus}
\end{figure}

Some energy spectra and edge states of the Hamiltonian (\ref{Eq:Ham2}) with $t_a=t_b$ are represented in Fig. \ref{Fig:Edgestateplus} for different values of $t_a$ (and $t_{ab}=t_{ba}=0$). There is an interesting symmetry between energy levels. To each energy eigenvalue $E$ for a given value of $t_a$ there exists an eigenvalue $-E$ if $t_a$ is replaced by $-t_a$. This is easily understood as follows. The simple transformation 
$a_{i\sigma}\rightarrow -a_{i\sigma}$, $b_{i\sigma}\rightarrow b_{i\sigma}$ changes the sign of both $H_{\mbox{\scriptsize SSH}}$ and 
$H_{\mbox{\scriptsize SO}}$, but leaves $H_{NNN}$ invariant. If, however, at the same time $t_a$ is replaced by $-t_a$, also $H_{NNN}$ changes sign.

To investigate the stability of the edge states, we have added a perturbation of the form
\begin{equation}\label{Eq:pertHamiltonian}
H_{p}= \sum_{i\sigma}U_\sigma\left(a_{i\sigma}^{\dagger}a_{i\sigma}^{\phantom{}}+b_{i\sigma}^{\dagger}b_{i\sigma}^{\phantom{}}\right)\, .
\end{equation}
This term breaks parity. We found that its addition quickly eliminates the edge states, which therefore appear to be ``protected by parity''. However, in contrast to chiral symmetry, which protects both the energy and the number of edge states (as long as the band gap is finite), this is no longer true in the present case. Fig.  \ref{Fig:Edgestateplus} exemplifies that for finite NNN hopping the levels of edge states move and even may merge with band states before the band gap closes.

%%%%%%%%%%%%%%%%%%%%%%%%%%%%%%%%%%%%%%%%%%%%%
\subsection{Broken parity, $CP$ invariance and a physical interpretation of the Zak phase}\label{sec:CDW}
\begin{figure}[h!]
	\centering
	\includegraphics[width=0.3\textwidth]{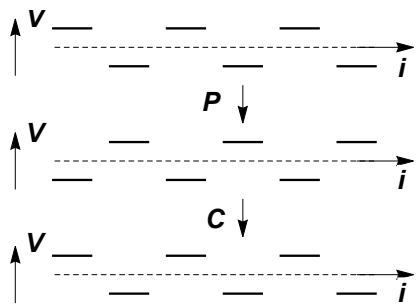}
	\caption{$CP$ symmetry of the alternating on-site potential. The $C$ and $P$ operations are up-down and right-left reflections, respectively.}
	\label{Fig:alternating}
\end{figure}

%%%%%%%%%%%%%%%%%%%%%%%%%%%%%%%%%%%%%%%%%%%%%
As an interlude, we consider the Hamiltonian $H_{\mbox{\scriptsize AA}}$, Eq. (\ref{Eq:ham_CDW}), for which both parity $P$ and charge conjugation $C$ are broken. However, as illustrated in Fig. \ref{Fig:alternating}, a parity operation followed by charge conjugation restores the original alternating potential. The model is $CP$-invariant.

In Fourier space the Hamiltonian reads
\begin{equation}\label{Eq:ham_CDW2}
H_{\mbox{\scriptsize AA}}=\sum_{k\sigma}\psi_{k\sigma}^\dag H_k\psi_{k\sigma}^{\phantom{}},
\end{equation}
where $\psi_{k\sigma}^\dag=(a_{k\sigma}^\dag, b_{k\sigma}^\dag)$ and
\begin{equation}\label{Eq:Bloch_CDW}
H_k=\left(\begin{array}{cc}\Delta& 1+e^{-ik}\\1+e^{ik}&-\Delta\end{array}\right).
\end{equation}
The eigenvalues are $\pm E_k$, where
\begin{equation}
E_k=\sqrt{4\cos^2\frac{k}{2}+\Delta^2}.
\end{equation}
At half filling, the system is insulating, with a gap $2\Delta$ separating valence and conduction bands. 
Normalized eigenstates of $H_k$ for negative eigenvalues $-E_k$ are
\begin{equation}\label{Eq:complex}
\left(\begin{array}{c}u_k^{(1)}\\ \\u_k^{(2)}\end{array}\right)=\frac{1}{\sqrt{2E_k}}
\left(\begin{array}{c}\sqrt{E_k-\Delta}\, e^{-i\frac{k}{4}}\\ \\-\sqrt{E_k+\Delta}\, e^{i\frac{k}{4}}\, .
\end{array}\right).
\end{equation}
The calculation of the Zak phase is greatly facilitated by the fact that the derivative of an even function is an odd function. Therefore only the derivatives of  the phase factors $e^{\pm ik/4}$ give finite contributions to the integral in Eq. (\ref{Eq:Berry}). We obtain (a factor of 2 comes from the spin)
\begin{equation}
\Phi=2i\sum_{\nu=1}^2\int_{-\pi}^{\pi}dk\, u_k^{(\nu)*}\partial_k u_k^{(\nu)}=-\frac{1}{2}\int_{-\pi}^\pi dk\, \frac{\Delta}{E_k}\, . 
\end{equation}
This expression can be written explicitly in terms of the elliptic integral $K$, but the important points are, on the one hand, that $\Phi$ is not quantized, in contrast to the case of the SSH Hamiltonian, on the other hand, that $\Phi$ is proportional to the charge polarization $P_{\mbox{\scriptsize ch}}$, as will be demonstrated now. 

The Hamiltonian (\ref{Eq:ham_CDW2}) is diagonalized by the Bogolyubov transformation
\begin{eqnarray}
a_{k\sigma}&=\cos \vartheta_k\, \alpha_{k\sigma}+e^{-i\frac{k}{2}}\sin\vartheta_k\, \beta_{k\sigma},\nonumber\\ 
b_{k\sigma}&=-e^{i\frac{k}{2}}\sin \vartheta_k\, \alpha_{k\sigma}+\cos\vartheta_k\, \beta_{k\sigma}\, .
\end{eqnarray}
With the choice
\begin{equation}
\tan 2\vartheta_k=-\frac{2}{\Delta}\cos\frac{k}{2}\, ,\quad \cos 2\vartheta_k=\frac{\Delta}{E_k}.
\end{equation}
the transformed Hamiltonian is
\begin{equation}
H_{\mbox{\scriptsize AA}}=\sum_{k\sigma}E_k\left(\alpha_{k\sigma}^\dag \alpha_{k\sigma}^{\phantom{}}
-\beta_{k\sigma}^\dag \beta_{k\sigma}^{\phantom{}}\right)\, .
\end{equation}

It is now straightforward to calculate the particle densities on the two sublattices and therefore also the polarization
\begin{equation}\label{Eq:prop}
P_{\mbox{\scriptsize ch}}:=\langle b_{i\sigma}^\dag b_{i\sigma}^{\phantom{}}\rangle-\langle a_{i\sigma}^\dag a_{i\sigma}^{\phantom{}}\rangle\, .
\end{equation}
With $\langle \beta_{k\sigma}^\dag \beta_{k\sigma}^{\phantom{}}\rangle=1$ and
$\langle \alpha_{k\sigma}^\dag \alpha_{k\sigma}^{\phantom{}}\rangle=0$ we obtain (in the thermodynamic limit)
\begin{equation}
P_{\mbox{\scriptsize ch}}=\frac{1}{2\pi}\int_{-\pi}^\pi dk\frac{\Delta}{E_k}=-\frac{\Phi}{\pi}.
\end{equation}
The amazing proportionality between the Zak phase $\Phi$ and the polarization $P_{\mbox{\scriptsize ch}}$ is the basis of the modern theory of polarization \cite{King-Smith_93, Resta_94}. It shows that the Zak phase can have a deep physical meaning even when it is not quantized. We notice that the proportionality (\ref{Eq:prop}) only holds for the complex representation chosen in Eq. (\ref{Eq:complex}), for which $\Phi$ vanishes in the limit $\Delta\rightarrow 0$.

%%%%%%%%%%%%%%%%%%%%%%%%%
\subsection{$CP$ symmetry for $t_a=-t_b$}\label{sec:CP}
%%%%%%%%%%%%%%%%%%%%%%%%%

We now return to the generalized SSH Hamiltonian (\ref{Eq:Ham2}) and concentrate on the case $t_a=-t_b$ and $t_{ab}=t_{ba}=0$. This is the Aubry-Andr\'e model in disguise (Appendix \ref{sec:CDW-SSH}), with CP symmetry. In fact, one readily shows that CP acts on the Bloch Hamiltonian $H_k$ as $CP=\sigma_0\otimes\sigma_y$, and that 
$CPH_k^*(CP)^{-1}=-H_k$, if $t_a=-t_b$. Such a system has two topological-insulator phases, ``protected by CP symmetry'' \cite{PhysRevB.90.245111}.

\begin{figure}[h!]
	\centering
	\includegraphics[width=0.5\textwidth]{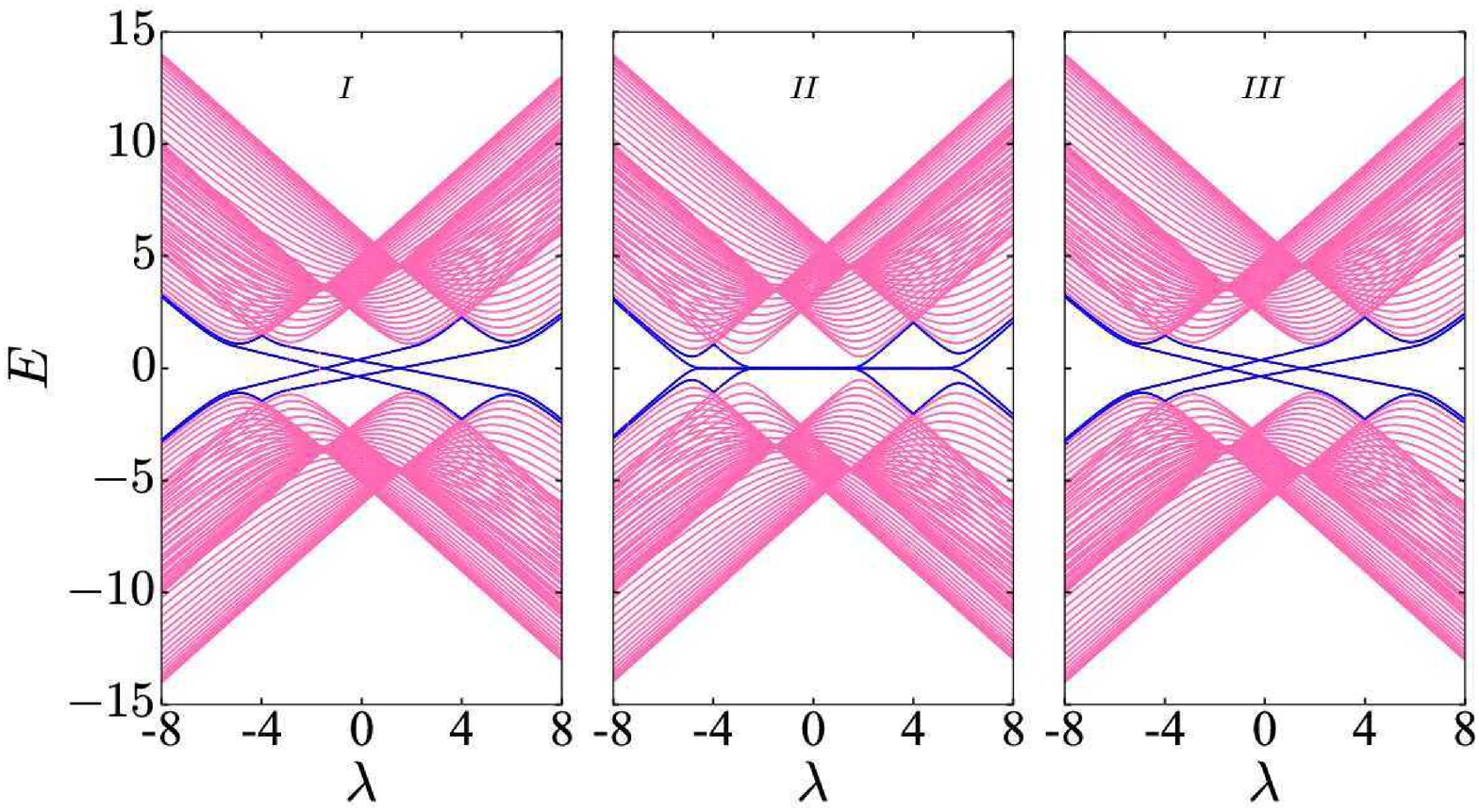}
	\includegraphics[width=0.5\textwidth]{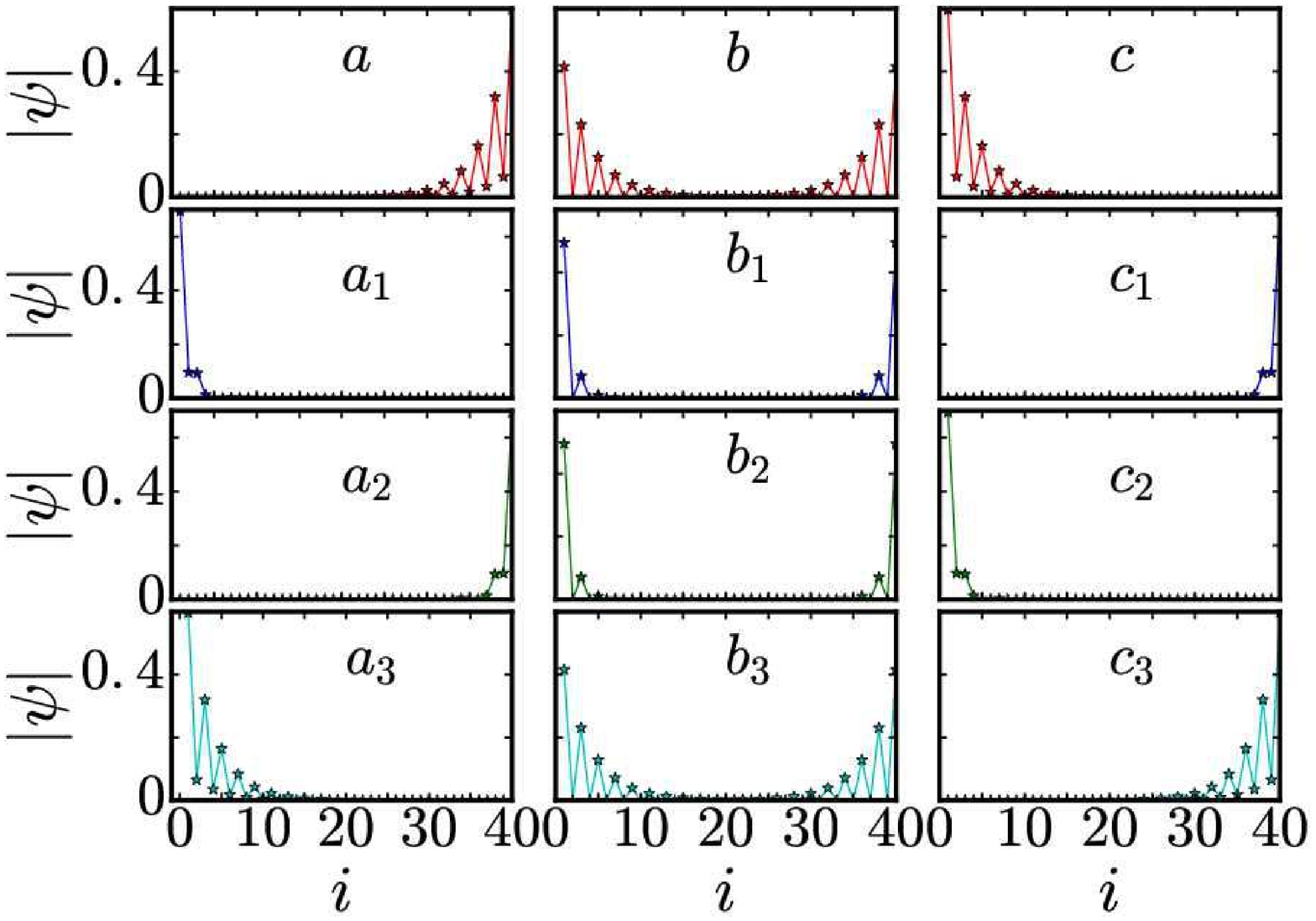}
	\caption{(Color online) Top row: the energy spectrum of a chain with $L=60$ sites versus $\lambda$, at $\lambda'=4$, and different strengths of second NN hoppings: (I) $t_a=-t_b=-0.5$, (II) $t_a=t_b=0$, and (III) $t_a=-t_b=0.5$. Bottom row: the probability amplitude of the edge states versus the system size L, at $\lambda=-1$ related to the energies $E=-0.54$, $-0.13$, $0.13$, and $0.54$, from up to down in left and right plots, and $E=0$ for the middle plot. }
	\label{Fig:Edgestate}
\end{figure}

Energy spectra and associated edge states for the CP-symmetric case ($t_a=-t_b$) are shown in Fig. \ref{Fig:Edgestate} for $t_a=-0.5$ (left) and $t_a=+0.5$ (right), in comparison to the case of chiral symmetry ($t_a=0$, middle). The results differ from those of the parity-symmetric case (Fig. \ref{Fig:Edgestateplus}) in two respects. On the one hand, in Fig. \ref{Fig:Edgestate} the spectra have particle-hole symmetry (not so in Fig. \ref{Fig:Edgestateplus}), on the other hand, the edge states of Fig. \ref{Fig:Edgestate} (for $t_a\neq 0$) are not reflection-symmetric, but rather concentrated on the left or right end of the chain (while they are symmetric in Fig. \ref{Fig:Edgestateplus}). Edge states no longer have zero energy for $t_a\neq 0$ because the chiral symmetry, which protects the zero-modes for $t_a=t_b=0$, is broken. Nevertheless, the CP symmetry protects the edge modes to some extent because if it is broken these states quickly disappear. We have verified this behavior explicitly by adding the perturbation (\ref{Eq:pertHamiltonian}), which breaks the CP symmetry.

% % % % % % % % % % % % % % % % % % % % % % % % %
\begin{figure}[h!]
	\centering
	\includegraphics[width=0.5\textwidth]{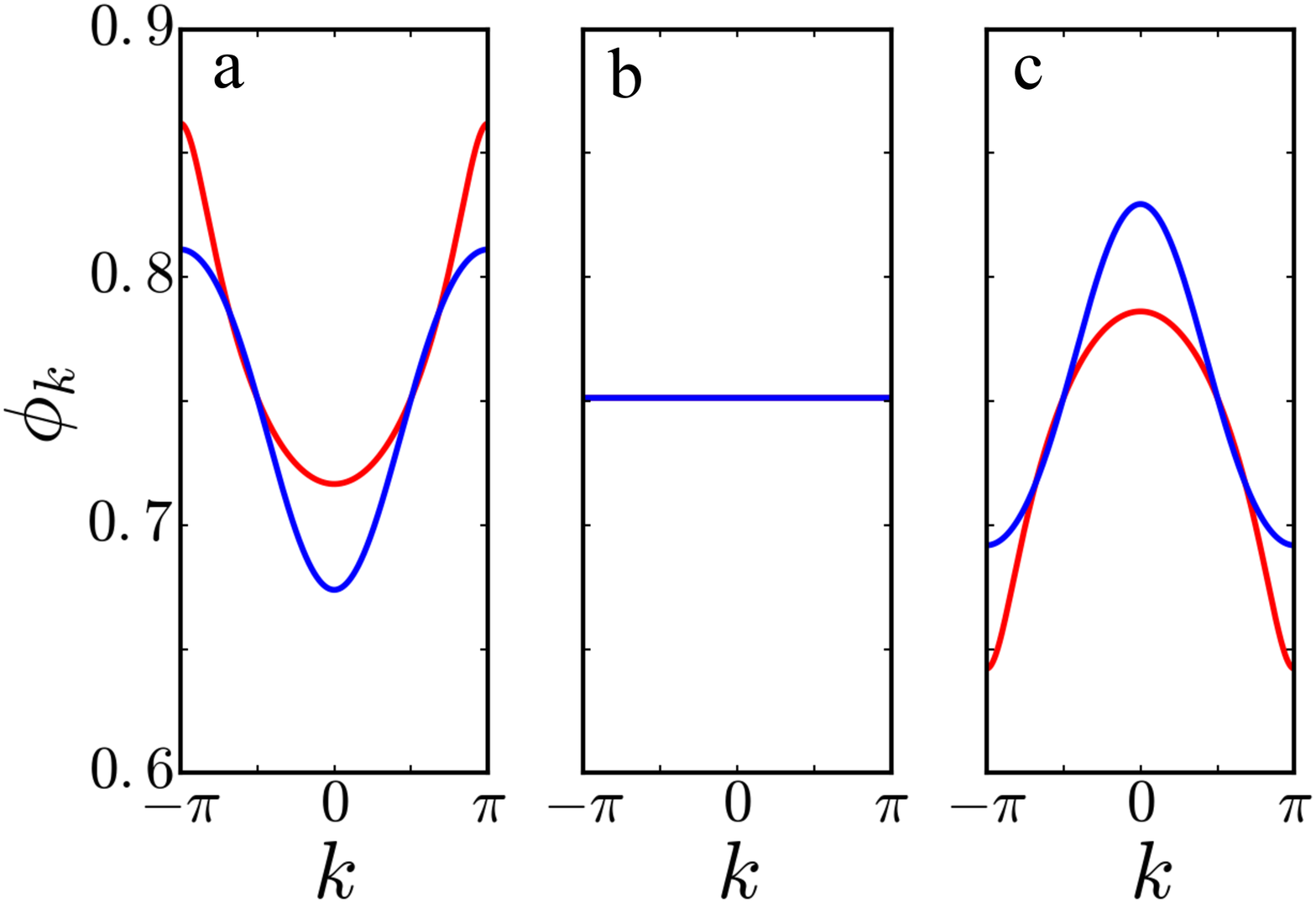}
	\includegraphics[width=0.5\textwidth]{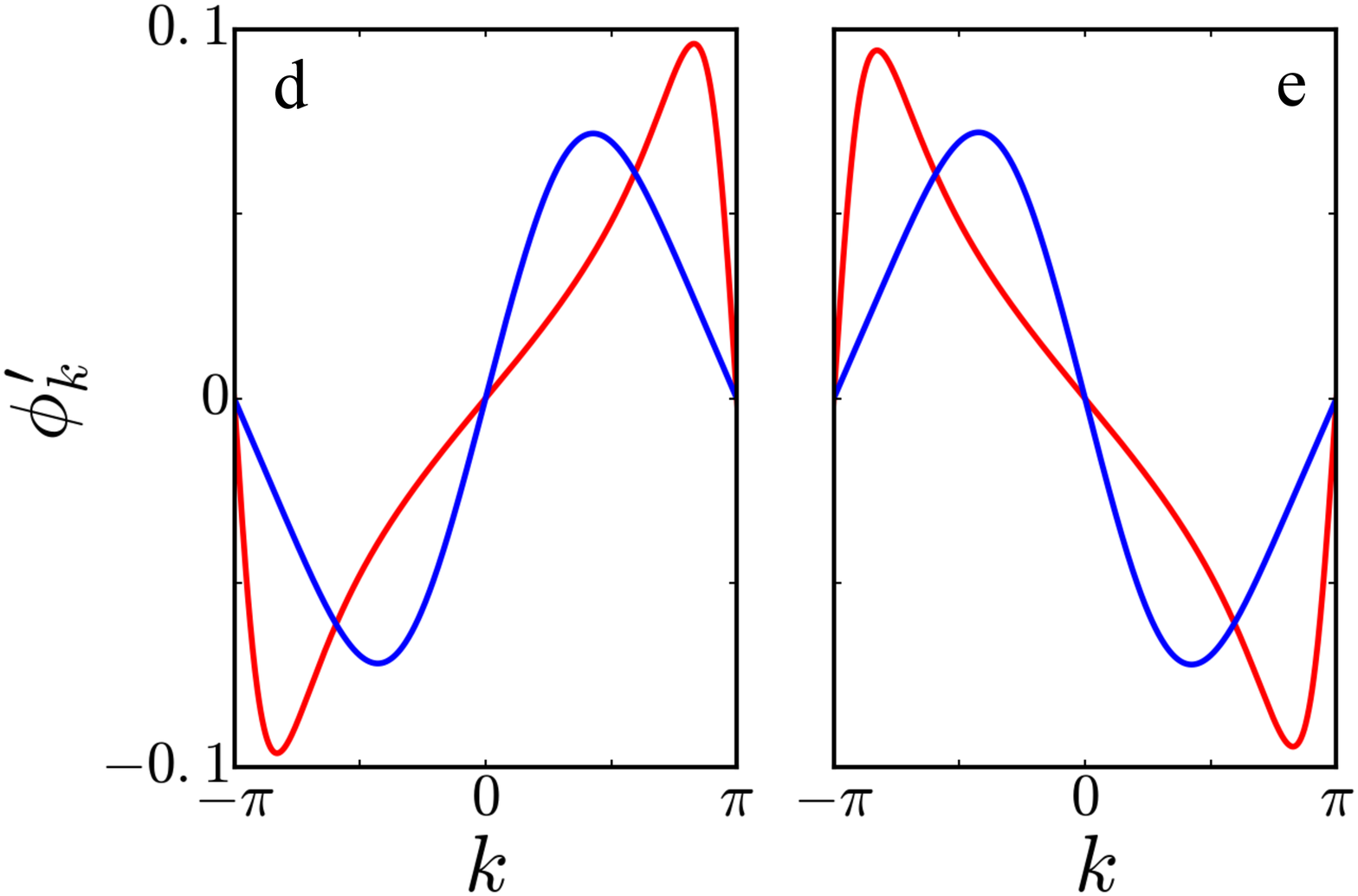}
	\caption{(Color online) Top: The ES of the SSH model with SOC in the presence of second NN hoppings, obtained by tracing the sublattice B out, for different $t_a$. (a) $t_a=-0.5$, (b) $t_a=0$, and (c) $t_a=0.5$. Bottom: the entanglement band velocity at (d) $t_a=-0.5$, and (e) $t_a=0.5$.}
	\label{Fig:ESBtrace}
\end{figure}

% % % % % % % % % % % % % % % % % % % % % % % % % %
An interesting aspect of Fig. \ref{Fig:Edgestate} is that, while the level spectrum is independent of the sign of $t_a$, the wave functions of the in-gap states are located at opposite edges. The entanglement spectrum can shed some light on the role of these two different cases. A useful cut is obtained by tracing out the $b$ sites, which leaves us with a chain of $a$ sites. The reduced correlation matrix in momentum space is 
\begin{equation}
G=
\begin{pmatrix}
\langle a_{k\uparrow}^{\dagger}a_{k\uparrow}^{\phantom{}}\rangle  & \langle a_{k\uparrow}^{\dagger}a_{k\downarrow}^{\phantom{}}\rangle  \\
\langle a_{k\downarrow}^{\dagger}a_{k\uparrow}^{\phantom{}}\rangle   &  \langle a_{k\downarrow}^{\dagger}a_{k\downarrow}^{\phantom{}}\rangle
\end{pmatrix}.
\end{equation}
Its eigenvalues $\eta_k^{(1)}$ and $\eta_k^{(2)}$ lead to the ES 
\begin{equation}
\phi_k^{(i)}=1-2\arctan\eta_k^{(i)}, \quad\quad\quad     i=1,2.
\label{Eq:ESBtrace}
\end{equation}
Fig. \ref{Fig:ESBtrace} shows that the ES depends sensitively on $t_a$. For vanishing $t_a$ the ES is dispersionless. We have verified that is true for all SPT phases of Fig. \ref{Fig:3rd-phase-diag}. For finite $t_a$ the ES is switched upside down if $t_a$ is replaced by $-t_a$. This behavior is well illustrated in terms of the {\it entanglement band velocity}, the group velocity of the ES (lower part of Fig. \ref{Fig:ESBtrace}), which changes sign if $t_a$ is replaced by $-t_a$. The entanglement band velocity is thus a valid quantity both for distinguishing CP from SPT phases and for telling the two CP phases from each other.

% % % % % % % % % % % % % % % % % % % % % % % % % % % % %
% % % % % % % % % % % % % % % % % % % % % % % % % % % % % % % % % % % %

We now show that also the Zak phase can be used for separating the various phases.
For simplicity we neglect both SOC and third-neighbor hopping. Thus the Bloch Hamiltonian is a 
$2\times 2$ matrix as in Eq. (\ref{Eq:Bloch_CDW}) with $\Delta$ replaced by $\omega_k=2t_a \cos k$ and with $1+\delta+(1-\delta)e^{\mp ik}$ as non-diagonal elements. We choose a complex representation as in Eq. (\ref{Eq:Bloch_CDW}) and find
\begin{equation}\label{eq:Zak}
\Phi=-\frac{1-\delta}{2}\int_{-\pi}^\pi dk\, \frac{\omega_k}{E_k}\frac{1-\delta+(1+\delta)\cos k}{E_k^2}\, ,
\end{equation}
where $E_k=\sqrt{\omega_k^2+2[1+\delta^2+(1-\delta^2)\cos k]}$. The Zak phase vanishes for $t_a\rightarrow 0$, as expected, but it also vanishes for $\delta\rightarrow \pm 1$, where the system becomes a two-leg ladder with hopping $\pm t_a$ along the legs and $2$ on the rungs. Moreover, $\Phi$ changes sign if $t_a$ is replaced by $-t_a$.

Is there a physical quantity which is proportional to $\Phi$, as in the CDW case? Proceeding as in Section \ref{sec:CDW} we can easily calculate correlation functions. The density is homogeneous, but the bond strengths $\langle a_{n\sigma}^\dag a_{n+1\sigma}^{\phantom{}}\rangle$ and $\langle b_{n\sigma}^\dag b_{n+1\sigma}^{\phantom{}}\rangle$ are different and depend on $t_a$. We find
\begin{equation}
\langle b_{n\sigma}^\dag b_{n+1\sigma}^{\phantom{}}\rangle-\langle a_{n\sigma}^\dag a_{n+1\sigma}^{\phantom{}}\rangle
=\frac{1}{2\pi}\int_{-\pi}^\pi dk\, \frac{\omega_k \cos k}{E_k}\, .
\end{equation}
 This quantity is somewhat different from the Zak phase, Eq. (\ref{eq:Zak}), but for small values of $t_a$ both expressions vary linearly with $t_a$ and tend to zero for $t_a\rightarrow 0$. A similar asymmetry is found if the ES, obtained by tracing out the $b$ sites, is compared with the ES calculated by tracing out the $a$ sites. For a fixed value of $t_a$ the corresponding spectra are precisely those of the left and right parts of Fig. \ref{Fig:ESBtrace}, respectively.

\section{Summary and concluding remarks}\label{sec:summary}

In this paper we have investigated a generalized SSH model, which embodies bond alternation, spin-orbit coupling and hopping beyond nearest-neighbor sites. Depending on its parameters the model produces a rich variety of features, partly topological, partly non-topological. Two main cases have been considered. On the one hand, we have studied the model where hopping is only allowed to occur between different sublattices (between odd and even sites). The Hamiltonian has then the symmetries of charge conjugation, time reversal and chirality. This system shows a rich phase diagram with the well-known properties of topological insulators, including quantized topological invariants, symmetry-protected edge states and topological phase transitions. On the other hand, we have studied the effects of next-nearest-neighbor hopping (hopping within sublattices), which destroys charge-conjugation (and therefore also chiral) symmetry. This system has a topologically trivial ground state. Nevertheless, some signatures of topologically non-trivial states can remain intact, such as integer winding numbers or robust edge states, especially if the additional hopping term is weak.  

We have used various tools to characterize the different ground states. Entanglement, a concept of central importance in quantum information theory, has been very useful in our study. Specific quantities such as entanglement spectrum, entanglement entropy and entanglement gap have been determined and shown to provide detailed informations about topological transitions in the chirally symmetric case, but also about the (non-topological) ground states in the absence of chiral symmetry. In the particular case where both charge-conjugation symmetry $C$ and parity $P$ are broken, but $CP$ is conserved (this happens if the two hopping parameters between next-nearest-neighbor sites have the same magnitude but different signs) we have introduced an ``entanglement band velocity'' to distinguish this case from the topologically non-trivial case, where both $C$ and $P$ are conserved.

The Zak phase (the Berry phase for a one-dimensional Bloch band) has been a versatile tool, both for determining the winding numbers of different topological states and as a measure for parity-symmetry breaking. We have used the fact that the Zak phase is not gauge invariant by choosing different specific gauges in these two cases. For topologically non-trivial states, a mixed real-complex representation for the eigenfunctions yields winding numbers in one-to-one correspondence with the numbers of edge states (bulk-boundary correspondence), while a fully complex representation is appropriate for describing states with broken parity, because in this case the Zak phase is proportional to a physical quantity measuring the amount of symmetry breaking (the polarization in the closely related Aubry-Andr\'e model).

The specific case of broken chirality and conserved parity (for non-alternating next-nearest-neighbor hopping) shows also interesting and partly unexpected features. The Zak phase remains quantized for gapped states, although these are topologically trivial. At the same time some topological transitions (occurring in the absence of next-nearest-neighbor hopping) are broadend into metallic regions, due to overlaps between conduction and valence bands.
 %%%%%%%%%%%%%%%%%%%%%%% References %%%%%%%%%%%%%%%%%%%%%%%%%%%%%%%%%%%%
\begin{acknowledgments}
The authors would like to thank Mahsa Seyedheydari for her help at the initial stage of this manuscript. NA thanks Henrik Johannesson for useful discussions.
\end{acknowledgments}
\appendix

\section{From AA to SSH}\label{sec:CDW-SSH}

Here we show that a simple unitary transformation can eliminate the on-site term of the Hamiltonian (\ref{Eq:ham_CDW}). We consider the case of periodic boundary conditions and introduce new fermionic operators $\alpha_{i\sigma}, \beta_{i\sigma}$ through the Bogolyubov transformation
\begin{eqnarray}
a_{i\sigma}&=\cos\vartheta\, \alpha_{i\sigma}+\sin\vartheta\, \beta_{i\sigma}\, ,\nonumber\\
b_{i\sigma}&=-\sin\vartheta\, \alpha_{i\sigma}+\cos\vartheta\, \beta_{i\sigma}\, .
\end{eqnarray}  
Inserting these equations into Eq. (\ref{Eq:ham_CDW}) and requiring that there be no alternating on-site term in the new representation, we obtain the condition $\tan\, \vartheta=\Delta$  (we put $t=1$). With $\cos\, \vartheta=\frac{1}{\tau}$,
$\sin\, \vartheta=\frac{\Delta}{\tau}$, where $\tau:=\sqrt{1+\Delta^2}$, we find
\begin{eqnarray}
H_{\mbox{\scriptsize AA}}&=&{\displaystyle\sum}_{i\sigma}\bigg\{\tau\left(\alpha_{i\sigma}^\dagger \beta_{i\sigma}^{\phantom{}}+\mbox{h.c.}\right)\nonumber\\
&&+\frac{1+\tau}{2\tau}\left(\alpha_{i+1\sigma}^\dagger \beta_{i\sigma}^{\phantom{}}+\mbox{h.c.}\right)\nonumber\\
&& -\frac{\Delta}{2\tau}\left(\alpha_{i\sigma}^\dagger \alpha_{i+1\sigma}^{\phantom{}}
 -\beta_{i\sigma}^\dagger \beta_{i+1\sigma}^{\phantom{}}+\mbox{h.c.}\right)\nonumber\\
&&+\frac{1-\tau}{\tau}\left(\alpha_{i\sigma}^\dagger \beta_{i+1\sigma}^{\phantom{}}+\mbox{h.c.}\right) \bigg\}\, .
\end{eqnarray}
The various terms represent nearest-neighbor hopping (lines 1 and 2), second-neighbor hopping with alternating sign (line 3) and third-neighbor hopping (only on half of the possible bonds).
%%%%%%%%%%%%%%%%%%%%%%%%%%%%%%
\bibliography{SSH-AAB}
\end{document}